\documentclass[12pt]{article}
\usepackage{mathrsfs}
\usepackage[T1]{fontenc}
\usepackage{mathpazo}
\usepackage{setspace}
\usepackage{amsfonts}
\usepackage{amssymb}
\usepackage{amsmath}
\usepackage{esint}                   % For closed integrals
\usepackage{epsfig}
\usepackage{latexsym}
\usepackage{color}
\usepackage{graphicx}
\usepackage{hyperref}
\usepackage{nicefrac}
\usepackage[latin1]{inputenc}
\usepackage{pstricks}
\usepackage{slashed}
\usepackage{multirow}
\usepackage{ulem}
\usepackage{comment}
\usepackage[nosort]{cite}
\usepackage{datetime}
\usepackage{tikz-cd}
\usepackage{float}

\usepackage{comment}
\usepackage{ytableau}

 \usetikzlibrary{decorations.pathreplacing}

 \usetikzlibrary{decorations.pathreplacing}

\def\hybrid{\topmargin -30pt    \oddsidemargin 0pt %%%%%%%%%%%%%% Archive-30pt
        \headheight 0pt \headsep 0pt
        \textwidth 6.25in       % A4 paper
        \textheight 9.5in       % A4 paper
        \marginparwidth .875in
        \parskip 5pt plus 1pt   \jot = 1.5ex}

\hybrid

\def\baselinestretch{1.2}

\catcode`\@=11

\def\marginnote#1{}
\newcount\hour
\newcount\minute
\newtoks\amorpm
\hour=\time\divide\hour by60
\minute=\time{\multiply\hour by60 \global\advance\minute by-\hour}
\edef\standardtime{{\ifnum\hour<12 \global\amorpm={am}%
        \else\global\amorpm={pm}\advance\hour by-12 \fi
        \ifnum\hour=0 \hour=12 \fi
        \number\hour:\ifnum\minute<10 0\fi\number\minute\the\amorpm}}
\edef\militarytime{\number\hour:\ifnum\minute<10 0\fi\number\minute}
\def\draftlabel#1{{\@bsphack\if@filesw {\let\thepage\relax
   \xdef\@gtempa{\write\@auxout{\string
      \newlabel{#1}{{\@currentlabel}{\thepage}}}}}\@gtempa
   \if@nobreak \ifvmode\nobreak\fi\fi\fi\@esphack}
        \gdef\@eqnlabel{#1}}
\def\@eqnlabel{}
\def\@vacuum{}
\def\draftmarginnote#1{\marginpar{\raggedright\scriptsize\tt#1}}

\def\draft{\oddsidemargin -.5truein
        \def\@oddfoot{\sl preliminary draft \hfil
        \rm\thepage\hfil\sl\today\quad\militarytime}
        \let\@evenfoot\@oddfoot \overfullrule 3pt
        \let\label=\draftlabel
        \let\marginnote=\draftmarginnote
   \def\@eqnnum{(\theequation)\rlap{\kern\marginparsep\tt\@eqnlabel}%
\global\let\@eqnlabel\@vacuum}  }

\def\draft2{
        \def\@oddfoot{\sl preliminary draft \hfil
        \rm\thepage\hfil\sl\today\quad\militarytime}
        \let\@evenfoot\@oddfoot \overfullrule 3pt
        \let\marginnote=\draftmarginnote
   \def\@eqnnum{(\theequation)\rlap{\kern\marginparsep\tt\@eqnlabel}%
\global\let\@eqnlabel\@vacuum}  }

\def\preprint{\twocolumn\sloppy\flushbottom\parindent 2em
        \leftmargini 2em\leftmarginv .5em\leftmarginvi .5em
        \oddsidemargin -.5in    \evensidemargin -.5in
        \columnsep .4in \footheight 0pt
        \textwidth 10.in        \topmargin  -.4in
        \headheight 12pt \topskip .4in
        \textheight 6.9in \footskip 0pt
        \def\@oddhead{\thepage\hfil\addtocounter{page}{1}\thepage}
        \let\@evenhead\@oddhead \def\@oddfoot{} \def\@evenfoot{} }

\def\numberbysection{\@addtoreset{equation}{section}
        \def\theequation{\thesection.\arabic{equation}}}

\def\underline#1{\relax\ifmmode\@@underline#1\else
        $\@@underline{\hbox{#1}}$\relax\fi}

\def\titlepage{\@restonecolfalse\if@twocolumn\@restonecoltrue\onecolumn
     \else \newpage \fi \thispagestyle{empty}\c@page\z@
        \def\thefootnote{\fnsymbol{footnote}} }

\def\endtitlepage{\if@restonecol\twocolumn \else \newpage \fi
        \def\thefootnote{\arabic{footnote}}
        \setcounter{footnote}{0}}  %\c@footnote\z@ }

\catcode`@=12
\relax

\def\figcap{\section*{Figure Captions\markboth
        {FIGURECAPTIONS}{FIGURECAPTIONS}}\list
        {Figure \arabic{enumi}:\hfill}{\settowidth\labelwidth{Figure
999:}
        \leftmargin\labelwidth
        \advance\leftmargin\labelsep\usecounter{enumi}}}
 \relax
\def\tablecap{\section*{Table Captions\markboth
        {TABLECAPTIONS}{TABLECAPTIONS}}\list
        {Table \arabic{enumi}:\hfill}{\settowidth\labelwidth{Table
999:}
        \leftmargin\labelwidth
        \advance\leftmargin\labelsep\usecounter{enumi}}}
 \relax
\def\reflist{\section*{References\markboth
        {REFLIST}{REFLIST}}\list
        {[\arabic{enumi}]\hfill}{\settowidth\labelwidth{[999]}
        \leftmargin\labelwidth
        \advance\leftmargin\labelsep\usecounter{enumi}}}
 \relax
\makeatletter
\newcounter{pubctr}
\def\publist{\@ifnextchar[{\@publist}{\@@publist}}
\def\@publist[#1]{\list
        {[\arabic{pubctr}]\hfill}{\settowidth\labelwidth{[999]}
        \leftmargin\labelwidth
        \advance\leftmargin\labelsep
        \@nmbrlisttrue\def\@listctr{pubctr}
        \setcounter{pubctr}{#1}\addtocounter{pubctr}{-1}}}
\def\@@publist{\list
        {[\arabic{pubctr}]\hfill}{\settowidth\labelwidth{[999]}
        \leftmargin\labelwidth
        \advance\leftmargin\labelsep
        \@nmbrlisttrue\def\@listctr{pubctr}}}
 \relax
\makeatother

\def\be{\begin{equation}}
\def\ee{\end{equation}}
\def\ba{\begin{eqnarray}}
\def\ea{\end{eqnarray}}

\def\del{\partial}

\def\bx {{\bf x}}
\def\bw {{\bf w}}

\def\r{\rho}
\def\a{\alpha}

\def\b{\beta}

\def\D{\Delta}

\def\m{\mu}

\def\l{\lambda}
\def\L{\Lambda}

\def\no{\noindent}

\def\qq{\qquad}

\def\IR{\relax{\rm I\kern-.18em R}}

\def\bx {{\bf x}}
\def\bw {{\bf w}}

\def\inv{^{\raise.0ex\hbox{${\scriptscriptstyle -}$}\kern-.05em 1}}

\def \ha {{\frac{1}{2}}}

\def \ov {\over}

\def\diag{{\rm diag}}

\newcommand{\bb}{\hskip -0.1cm}
\newcommand{\hb}{\hskip -0.05cm}
\newcommand{\ssb}{\hskip -0.03cm}

\def\tr{\textrm{Tr}}

\begin{document}
%\draft2

%\renewcommand{\theequation}{\arabic{equation}}
%\renewcommand{\theequation}{\thesection.\arabic{equation}}

\renewcommand{\theequation}{\thesection.\arabic{equation}}
\csname @addtoreset\endcsname{equation}{section}

\begin{titlepage}
\begin{center}

\renewcommand*{\thefootnote}{\arabic{footnote}}

\phantom{xx}
\vskip 0.5in

{\large {\bf Adjoint ferromagnets}} 
%and conjugation symmetry breaking

\vskip 0.5in

 {\bf  Joaqu\'in L\'opez-Su\'arez }$^{1}$, \hskip .2 cm  {\bf Alexios P. Polychronakos}$^2$, \hskip .15cm
{\bf Konstantinos Sfetsos}$^{1}$

\vskip 0.15in

${}^1\!$
Department of Nuclear and Particle Physics, \\
Faculty of Physics, National and Kapodistrian University of Athens, \\
Athens 15784, Greece\\
{\footnotesize{\tt  jlopezs@phys.uoa.gr,\hskip .15cm ksfetsos@phys.uoa.gr}}

\vskip .15in

${}^2\!$ Physics Department, the City College of New York, NY 10031, USA \\

%\vskip .2 cm 

%\vskip .2 cm
The Graduate Center of the CUNY, New York, NY 10016, USA\\
{\footnotesize{\tt apolychronakos@ccny.cuny.edu}}; {\footnotesize{\tt apolychronakos@gc.cuny.edu}}

\vskip .3in
\today

\vskip .2in

\end{center}

\vskip .2in

\centerline{\bf Abstract}

\no
We derive the phase structure and thermodynamics of ferromagnets consisting of
elementary magnets carrying the adjoint representation of $SU(N)$ and coupled through
two-body quadratic interactions. Such systems have a continuous $SU(N)$ symmetry as well
as a discrete conjugation symmetry. We uncover a rich spectrum of phases and transitions, involving
a paramagnetic and two distinct ferromagnetic phases that can coexist as stable and metastable states in different combinations
over a range of temperatures. The ferromagnetic phases break $SU(N)$ invariance in various channels,
leading to spontaneous magnetization.
Interestingly,  the conjugation symmetry also breaks over a range of temperatures and group ranks $N$, providing a
realization of a spontaneously broken discrete symmetry.
\vskip .4in

\vfill

\end{titlepage}
\vfill
\eject

%\def\baselinestretch{1.2}
%\baselineskip 10 pt
%\noindent

%\tableofcontents

\def\baselinestretch{1.2}
\baselineskip 20 pt

\newcommand{\eqn}[1]{(\ref{#1})}

\tableofcontents

%%%%%%%%%%%%%
\section{Introduction}

Magnetism with $SU(N)$ degrees of freedom has been of increasing relevance, both in theoretical and experimental settings. 
Magnetic systems with $SU(N)$ symmetry have been considered in the context of ultracold atoms
\cite{Dud,Ghu,Gor,Zha,Mag,Cap,Mukherjee:2024ffz}, spin chains \cite{Aff,Pola},
interacting atoms on lattice cites \cite{KT,BSL,RoLa,YSMOF,TK,Totsuka,TK2},
and were also studied in the presence of external $SU(N)$ magnetic fields \cite{DY,YM,HM}.

\no
The thermodynamics and phase transitions of $SU(N)$ ferromagnets, in particular, have proven extremely rich.
In our recent work \cite{PSferro, PSlargeN, PSferroN, PShigher,PScubic} we studied ferromagnets
with each elementary magnet ("atom") carrying $SU(N)$ degrees of freedom and two- or higher-body interactions and uncovered an intricate and
nontrivial phase structure with qualitatively new features compared to standard $SU(2)$ ferromagnets,
even in situations where the only relevant parameter is the rank $N$ of the group.
We established the existence of several critical temperatures (as opposed to the single Curie temperature
for $SU(2)$), metastability regimes, hysteresis phenomena, and spontaneous breaking of the global $SU(N)$ symmetry
into various channels. We also uncovered zero-temperature (quantum) phase transitions in the presence of three-body interactions.
New critical phenomena emerge as the rank of the group becomes large, with a triple critical point appearing
in an appropriate scaling limit.

\no
The $SU(N)$ representation carried by each atom is a defining element of the ferromagnet.
The fundamental $N$-dimensional representation corresponds to the $N$ internal states fully mixing under the action
of the symmetry group. Higher representations correspond to reduced symmetry, in which interactions are still
$SU(N)$-invariant, but the atom states are not arbitrarily mixed under the action of the symmetry group. In \cite{PShigher} we
developed the formalism for dealing with a general atom representation $\chi$, and studied in detail the cases of the symmetric and antisymmetric
irreducible representations (irreps), revealing the existence of additional phases and spontaneous symmetry breaking patterns.

\no
In this paper we perform the analysis of the ferromagnet with atoms in the adjoint representation by using the general formalism of \cite{PShigher}.
This is a particularly interesting case.
The adjoint irrep is self-conjugate, and consequently the ferromagnet has an additional discrete conjugation invariance symmetry.
In this situation, one would expect
a lower propensity to spontaneous $SU(N)$ symmetry breaking, as most breaking channels have a preferred magnetization and would
break the conjugation symmetry. Nevertheless, our analysis reveals that symmetry breaking phases with nontrivial magnetization do arise,
all of them breaking $SU(N)$ down to smaller subgroups, and some of them also breaking conjugation invariance. 
The phase structure is, if anything, richer than the
one for the previously studied atom irreps and depends nontrivially on the rank $N$, with various critical temperatures emerging and
rearranging as $N$ increases, presenting a large set of phase transition scenaria. To probe the sensitivity of the model to relatively minor
variants of its dynamics, we also study the case of the reducible fundamental-antifundamental atom representation (differing from the adjoint
by the presence of an additional singlet state) and show that its phase structure is qualitatively similar to the one of the adjoint.

\no
The organization of the paper is as follows:
In section \ref{model} we review the model of atoms with a general atom representation $\chi$, lay out
the essential group theory facts required for its analysis, and derive its thermodynamic equilibrium and stability conditions.
In section \ref{adjoint} we specialize to the case of the adjoint atom irrep, study its thermodynamics and stability conditions, and determine
the structure of its possible stable phases. In section \ref{detailed} we perform the analysis of the stable phases for various values of the
group rank $N$, derive their critical temperatures, and determine their symmetry breaking patterns and cascade of phase transitions.
We also perform the analysis of the reducible fundamental-atifundamental representation and compare the results to those for the adjoint.
Finally, in section \ref{conclusions} we present our conclusions. A detailed analysis of unstable solutions and some group theory results
are given in the appendices.

  %%%%%%%%
  \section{ The model and the thermodynamic limit}\label{model}
 
 We consider $n \gg 1$ atoms at fixed positions, each carrying an irrep $\chi$ of $SU(N)$ and interacting with ferromagnetic two-body
 interactions. The analysis of \cite{PSferro, PShigher} shows that, in the mean field approach, the ferromagnet can be described in terms of
 an effective coupling $c$ and states transforming as irreps of the global $SU(N)$ symmetry of the system, and we refer the reader to these
 papers for the details of the derivation. The relevant mathematical quantities are the number of times each irrep of the global $SU(N)$ is contained in the
 Hilbert space and the quadratic Casimirs of these irreps.
 
 \no
 Irreps of $SU(N)$ can be represented in terms of decreasing nonnegative integers $k_i$, $i=1,2,\dots,N$, related to Young tableau (YT) row lengths as
 $\ell_i = k_i -k_N +i-N$ \cite{PScompo}.
 The multiplicity of irreps $d_{n,{\bf k}}$ in the decomposition of the direct product of $n$ representations $\chi$ of $SU(N)$
 with character $\chi({\bf z})$ is given by 
 \be
 \sum_{\bf k} d_{n;{\bf k}} z_1^{k_1}\dots z_N^{k_N} = \D({\bf z}) \chi^n({\bf z})\ ,
 \ee
with $z_i$ auxiliary variables corresponding to the eigenvalues of the fundamental group element $U = \diag\{{\bf z}\}$ (boldface $\bf k$ stands for the collection $\{k_i\}$ and similarly for $\bf z$) and $\D({\bf z})$ being the Vandermonde determinant.
 Then $d_{n;{\bf k}}$ obtains from a multiple integration in the complex $\bf z$-plane around the origin as 
 \be
 d_{n;{\bf k}} = {1\ov (2 \pi i)^N} \prod_{i=1}^N   \oint_{_C}  {d z_i\ov z_i^{k_i+1}} \D({\bf z}) \chi^n({\bf z})\ .
 \ee
The partition function of the ferromagnet with atoms carrying the representation $\chi$, in the presence of magnetic fields $B_i$ along the Cartan subalgebra of $SU(N)$, is given by
\be
Z= \sum_{\text{states}} e^{-\beta H} =
\sum_{<{\bf k}>} d_{n;{\bf k}}\, e^{{\beta c\over n} C^{(2)} ({\bf k})}\,
\tr_{\bf k} \exp\Bigl( \beta \sum_{j=1}^N B_j H_j  \Bigr)\ ,
\label{Zo}\ee
$<$${\bf k}$$>$ representing non-negative decreasing integers and $C^{(2)} ({\bf k})$ the quadratic Casimir
of the irrep $\bf k$.
We are interested in the thermodynamic limit in which $n\gg 1$. Setting 
\be
k_i = n x_i\ 
\ee 
the $x_i$ are of order 1 in the $n\gg 1$ limit and constitute the magnetization parameters of the ferromagnet \cite{PSferro}.
Keeping only leading orders in $n$, we obtain \cite{PShigher}
 \be
Z= {n^N\ov (2 \pi i)^N}  \prod_{i=1}^N \int dx_i   \oint  {  dz_i\ov z_i} e^{-n \b   F({\bf x},{\bf z})}\ ,
\label{Z}\ee 
where 
\be
\label{ufhju1}
F({\bf x},{\bf z})=  \sum_{i=1}^N \Big( T  x_i \ln z_i- {c x_i^2\ov 2} - B_i x_i\Big) -T \ln \chi({\bf z})\ ,
\ee
is the free energy per atom.

\subsection{Equilibrium equations}

For $n\gg 1$, the integral in \eqn{Z} is dominated by the saddle point of $F({\bf x},{\bf z})$. In the saddle-point approximation,
$F({\bf x},{\bf z})$ has to be stationary both in $z_i$ and $x_i$, leading to the equilibrium conditions
\be
\label{eqsfxz}
\begin{split}
& {\del F\ov \del z_i}= T\bigg( {x_i\ov z_i}  -  {\del_{z_i} \chi \ov \chi}\bigg) =0\ ,
\\
& {\del F\ov \del x_i}= T \ln z_i -c x_i - B_i  =0\ .
\end{split}
\ee
Note that even though the $z_i$ are in principle complex, reality of $x_i$ implies that the equilibrium $z_i$ are real and positive.
Hence, we define new variables $w_i$ and $\l$ as 
\be
 z_i = e^{w_i}\ ,\qq  \lambda =  \ln \chi\  ,
 \label{zl}
 \ee
 in terms of which the equilibrium equations for $\bf w$ and $\bf x$ become
 \be
 \label{neweq}
x_i =   {\del \l \ov \del w_i} ~, \quad T\, w_i  =  c x_i + B_i \ .
\ee
The free energy \eqn{ufhju1} is a function of both sets of variables $x_i$ and $w_i$.
Enforcing the equilibrium equation for $w_i$ gives the specific free energy in terms of the magnetization parameters $x_i$,
\be
F(\bx) = \sum_{i=1}^N \left( T x_i w_i (\bx) -{ c \over 2}x_i^2 -  B_i x_i \right) -T \lambda(\bw(\bx))\ ,
\label{Sx}
\ee
$w_i (\bf x)$ being the solution of the first equation in \eqn{neweq}. This expression is useful for the stability analysis around equilibrium
configurations, as it is valid for arbitrary values of $x_i$. Enforcing also the equilibrium equation for $x_i$,
the expression for $F(\bf x)$ simplifies to
\be
\label{fbx}
F(\bx) = \sum_{i=1}^N {c \over 2}x_i^2 -T \lambda\big(c\bx/T+{\bf B}/T\big) \ .
\ee
In fact, the above expression, taken off equilibrium, correctly reproduces both the equilibrium and the stability conditions for the system derived below.

\no
For an irreducible representation, the character $\chi$ is a homogeneous polynomial in the $z_i$ with degree of homogeneity $n_\chi$
the number of boxes in the YT of the irrep $\chi$. Summing the first of \eqn{neweq} over $i$, the right hand side gives
the degree of homogeneity of $\chi$ and we obtain
\be
\sum_{i=1}^N x_i = \sum_{i=1}^N  {\del \lambda \over \del w_i} = n_\chi\ .
\label{sx}
\ee
This is a {\it constraint} on the variables $x_i$, necessary for \eqn{neweq} to have a solution for the $w_i$. If it is
satisfied, the $w_i$ are only determined up to a common additive constant.

\no
We further note that the character $\chi ({\bf z})$ is defined up to the transformation
\be
\chi ({\bf z})\to (z_1 \cdots z_N )^m\, \chi({\bf z})\ , \qq m \in \mathbb{Z}\ ,
\ee
which changes the $U(1)$ charge but preserves the $SU(N)$ part of the irrep. From \eqn{neweq}, the equilibrium $x_i$ change as
$x_i \to x_i +m$ under this transformation, yielding the same irrep for the state, and the free energy \eqn{fbx} changes by a trivial constant.

\subsection{Stability}

We are interested in solutions of the equilibrium equations that are perturbatively stable.  
The analysis of \cite{PShigher} led to the stability criterion that the following two matrices should be positive semi-definite:
\be
\label{Lcond}
\L \geqslant 0\ ,\qq  T\, \mathbb{1} - c \L\geqslant 0\ , 
\ee 
with
\be
\Lambda_{ij} = \left.{\del^2 \lambda \over \del w_i \del w_j}\right|_{{\bf w} = {(\ssb c {\bf x} +{\bf B}\ssb)\ssb/\hb T}}
= \left.{1\over \chi}{\del^2 \chi \over \del w_i \del w_j}\right|_{{\bf w} = {(\ssb c {\bf x} +{\bf B}\ssb)\ssb/\hb T}}
- x_i \, x_j\ .
\label{Lx}
\ee
The subscript indicates that after the derivatives are taken we use the equilibrium equations.

%%%%%%%%
 \section{ The adjoint representation }\label{adjoint}
 
 We focus on the ferromagnet with atoms in the adjoint representation in the absence of external fields. In what follows,
 we set the magnetic fields  $B_i$ to zero.
 
 \no
 The YT for the adjoint representation is conveniently presented by using the notation 
 %%%
\begin{comment}
\be
\begin{array}{c}
\begin{ytableau}
\none & \boldsymbol{\bullet}
\end{ytableau}
\\[-0.4em]  % This works inside array
\quad\ \ \vdots
\\ [0.1 em] 
\begin{ytableau}
\none & \boldsymbol{\bullet}
\end{ytableau}
\end{array}
\ee
\end{comment}
%%%
\be
s~ \text{antiboxes} \left \{\hskip -.4 cm
\begin{array}{c}
\scalebox{.9}{%
\begin{ytableau}
\none & \boldsymbol{\bullet}
\end{ytableau}
}
\\[-0.4em]
\quad\  \vdots
\\[0.1em]
\scalebox{0.9}{%
\begin{ytableau}
\none & \boldsymbol{\bullet}
\end{ytableau}
}
\end{array}
\right. 
\hskip .35 cm = \left.
 \begin{array}{c}
\scalebox{.9}{%
\begin{ytableau}
\none &
\end{ytableau}
}
\\[-0.4em]
\quad\  \vdots
\\[0.1em]
\scalebox{0.9}{%
\begin{ytableau}
\none & 
\end{ytableau}
}
\end{array}
\right\} N\!-\! s~\text{boxes}\ .
\ee
Then, the YT for the adjoint representation is
\be
\label{ytadjoint}
\scalebox{.9}{%
\begin{ytableau}
\none & \boldsymbol{\bullet} & 
\end{ytableau}
}\ \ = \hskip -1 cm\left.
 \begin{array}{c}
\scalebox{.9}{%
\begin{ytableau}
\none \hskip .6 cm & & 
\end{ytableau}
}
\\[-0.4em]
\quad\  \vdots
\\[0.1em]
\scalebox{0.9}{%
\begin{ytableau}
\none & 
\end{ytableau}
}
\end{array}
\right\} N~\text{boxes}\ .
\ee
%The YT is \hskip -.17 cm b
%{\tiny \begin{ytableau}
%\none   &  \boldsymbol{ \bullet} & 
%\end{ytableau}} . 
The character in terms of the variables $z_i$ is \cite{PScompo}
\be
\label{hjk19}
\chi({\bf z})=\Big(\sum_{i=1}^N  z_i \Big) \Big(\sum_{i=1}^N  z_i^{-1} \Big) -1 \ .
\ee
This corresponds to the box-antibox representation, in which the total number of boxes is $n_\text{adj} =0$ (in the
representation in terms of boxes we would have $n_\text{adj} =N$). Consequently, the constraint  \eqn{sx} becomes
  \be
\label{zxad}
\sum_{i=1}^N  x_i = 0 \ .
  \ee
In the box-antibox representation, we can directly interpret $n x_i$ as the number of boxes (for $x_i>0$) or antiboxes (for $x_i <0$)
in row $i$ of the YT of the configuration $\bf x$. (The box-only representation corresponds to row lengths $\ell_i = n(x_i - x_N )$ to leading order in $n$.)
The free energy is invariant under the transformation $x_i \to -x_i$, which corresponds to turning the irrep
parametrized by $x_i$ to its conjugate. This conjugation invariance is a corollary of the fact that the adjoint irrep
of the atoms is self-conjugate.

\no
To efficiently analyze the equilibrium conditions, we define new 
%rescaled variables
%\be
%\label{jfn22}
%e^{\a_i} = \sqrt{\sum_{j=1}^N  z_j^{-1}  \ov \sum_{j=1}^N  z_j}\,   z_i \, , \quad  i=1,2,\dots, N\ ,
%\ee
variables 
\be
\label{jfn22}
\a_i = w_i + \m \ , \qq \m =  \ha \ln { \sum_{j=1}^N  z_j^{-1}  \ov \sum_{j=1}^N  z_j}\ ,
\ee
($\m$ absorbs the arbitrary constant in the definition of $w_i$). The $\a_i$ are subject to the constraints
\be
\label{raai}
 \sum_{i=1}^N e^{\a_i} = \sum_{i=1}^N e^{-\a_i} = \r \ , ~~~ %= \sqrt{\chi({\bf z}) +1} 
 \r^2 \equiv \sum_{i,j=1}^N  z_i\, z_j^{-1} = \chi(\bf z)+1\ .
\ee
Clearly $\r \geqslant N$. Then, the first of \eqn{neweq} gives 
\be
\label{xira}
x_i = {2\r\ov \r^2-1} \sinh \a_i \ ,\qq i=1,2,\dots , N\ ,
\ee 
which, due to the constraints \eqn{raai}, implies \eqn{zxad},
in agreement with \eqn{sx}. The second equation in \eqn{neweq} becomes 
\be
\label{eqq1}
\a_i- {T_0(\r)\ov T} \sinh \a_i = \m , \qq i=1,2,\dots , N\ ,
\ee
where
\be
\label{tor}
T_0(\r)= {2\r\ov \r^2-1}\, c\ ,\qq 0<T_0(\r)<  T_0(N) = {2 N\ov N^2-1}\, c \equiv T_0\ .
\ee
$T_0$ denotes the maximal value of $T_0(\r)$, in terms of which $c$ is expressed as 
\be
\label{cadj}
c = {N^2-1\ov 2 N }\, T_0\ .
\ee
Similarly to previous work \cite{PSferro, PShigher}, $T_0$ will turn out to be the temperature below which the singlet becomes unstable.
It is interesting that $c$ can be written in the general group theoretical form \eqn{cggrp} which also applies
to the fundamental, symmetric, and antisymmetric atom representations studied in  \cite{PShigher}.

\begin{figure} [h]
\begin{center}
\includegraphics[height= 5 cm, angle=0]{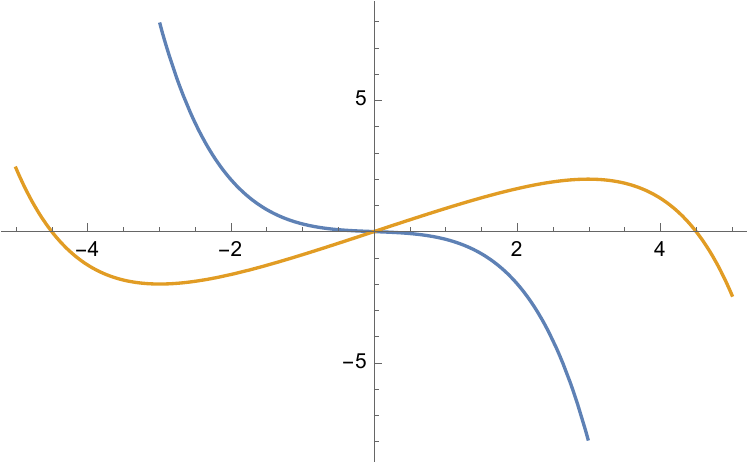}
\end{center}
\vskip -.6 cm
\caption{\small{Plots of the left hand side of \eqn{eqq1} for $T<T_0(\r)$ (blue)  and for $T>T_0(\r)$ (yellow).}}  
\label{adj1}
\end{figure}

\no
Equations \eqn{eqq1} and the two constraints \eqn{raai} determine the $\a_i$ and the two Lagrange multiplier-like variables $\r$ and $\m$.
For fixed $\r$ and $\m$, all $\a_i$ obey the same equation.
This equation can have either one or three solutions, depending on whether the ratio ${T_0(\r)/T}$ is larger or smaller 
than unity (see fig. \ref{adj1}).
%%%%%%
%%%%%%
However, the single solution case is  acceptable only for $\a_i=0$ (corresponding to $x_i =0$, $\r=N$) since otherwise 
 the constraint \eqn{zxad} cannot be satisfied. This corresponds to the singlet state with no spontaneous magnetization. 
 
\no
In the three solution case, not all of the $\a_i$ can have the same sign, otherwise the constraint \eqn{zxad} cannot be satisfied. 
Since the state is invariant under permutations of the $\a_i$, we may choose the ordering of the solutions
\be
\label{ordd}
\a_3<\a_2<\a_1\ ,\qquad \a_3<0\, ,\quad \a_1>0\ .
\ee
We will denote by $p_1$ the number of $\a_i$ corresponding to the solution $\a_1$ 
and similarly for the solutions $\a_2$ and $\a_3$. The $p_i$ clearly obey 
\be
\label{u11}
p_1+p_2+p_3= N\ ,\qq p_{1,2,3} =0,1,\dots , N-1\ .
\ee
Note that this condition for $N=2$ has as only solution (without loss of generality) $p_1=p_3=1,p_2=0$. The case of
$SU(2)$ will be treated separately.
%, corresponding to the one row YT with two boxes.

\no
From here on, we focus to $N\geqslant 3$. Then the constraint \eqn{zxad} becomes 
\be
\label{u14}
p_1 \sinh \a_1 +p_2 \sinh \a_2 +p_3 \sinh \a_3 =0 \ .
\ee
Finally,  \eqn{raai} implies
\be
\label{u12}
\r = p_1 e^{\a_1} +p_2 e^{\a_2} +p_3 e^{\a_3} = p_1 e^{-\a_1} +p_2 e^{-\a_2} +p_3 e^{-\a_3}\ .
\ee

\no
This solution corresponds to a YT with $p_1+p_2$ rows,
the first $p_1$ of them of equal length $\ell_1 =(x_1-x_3)n$ and the remaining $p_2$ of equal length $\ell_2 =(x_1-x_3)n$. 
%Using \eqn{xira} and \eqn{u13w} we obtain
That is,
\begin{equation}
\begin{tikzpicture}[baseline=(current bounding box.center)]
\node (yt) {
\scalebox{.7}{ \begin{ytableau}
\none   &  & & \none[\cdots] & \none[\cdots] & & & & \none[\cdots] & \none[\cdots] & \\
\none & \none[\vdots] & \none & \none & \none & \none & \none & \none & \none & \none & \none
\\
\none   &  & & \none[\cdots] & \none[\cdots] & & & & \none[\cdots] & \none[\cdots] & \\
\none   &  & & \none & \none & \\
\none   & \none[\vdots] & \none & \none & \none & \none
\\
\none   &  & & \none[\cdots] & \none[\cdots] &
\end{ytableau}
}
};
% p_1 brace
\draw[decorate,decoration={brace,amplitude=5pt}]
    ([xshift=10pt]yt.west) -- ([xshift=10pt]yt.north west)
node[midway,xshift=-0.6cm] {$p_1$};
    % p_1 brace
\draw[decorate,decoration={brace,amplitude=5pt}]
([xshift=10pt]yt.south west) -- ([xshift=10pt]yt.west)
node[midway,xshift=-0.6cm] {$p_2$};
% l_1 brace
\draw[decorate,decoration={brace,mirror,amplitude=5pt}]
    (yt.north east) -- ([xshift=15pt]yt.north west)
node[midway,yshift=0.6cm] {$\ell_1$};
    % l_2 brace
\draw[decorate,decoration={brace,mirror,amplitude=5pt}]
    ([xshift=15pt]yt.south west) -- ([xshift=15pt]yt.south)
node[midway,yshift=-0.6cm] {$\ell_2$};
\end{tikzpicture}\ 
\label{genn}
\end{equation}

\no
To solve the equations and determine $\a_{1,2,3}$, we first choose $p_i$ obeying \eqn{u11}. Then, assuming $p_3\neq 0$,
 we may use \eqn{u14} to solve for $\a_3$ in terms of $\a_1$ and $\a_2$
\be
\a_3 =-\ln\Big( f+ \sqrt{ f^2+1}\Big)\ ,\qq  f= {p_1\ov p_3} \sinh \a_1  +{p_2\ov p_3} \sinh \a_2\ 
\ee
and then determine the latter from the system of transcendental equations
\be
\begin{split}
\label{u13w}
& \a_1- {T_0(\r)\ov T} \sinh \a_1 = \a_2- {T_0(\r)\ov T} \sinh \a_2\ ,
\\
& \a_1- {T_0(\r)\ov T} \sinh \a_1 = \a_3- {T_0(\r)\ov T} \sinh \a_3\ , 
\end{split}
\ee
where $T_0(\r)$ in \eqn{tor}   is found using \eqn{u12}.
Hence, \eqn{u13w} is a system with two unknowns $\a_1$ and $\a_2$, which can be chosen as in \eqn{ordd}.
%\no
%Note that the space of solutions for $\a_{1,2,3}$ is invariant under flipping simultaneously their signs and also under an interchange of any two of them 
% so that we may restrict to those satisfying \eqn{ordd}.
The free energy at equilibrium is 
\be
\label{freeF}
\begin{split}
 F(T)  & =T_0 {\r^2 (N^2-1)\ov N (\r^2-1)^2} \sum_{i=1}^N \sinh^2 \a_i - T \ln (\r^2-1)
\\
&  = T_0\,  {\r^2 (N^2-1)\ov N (\r^2-1)^2} (p_1 \sinh^2 \a_1 + p_2 \sinh^2 \a_2  + p_3 \sinh^2 \a_3 )- T\, \ln (\r^2-1)\ ,
\end{split}
\ee
where we used the equilibrium equation \eqn{eqq1} and the constraint \eqn{zxad} and 
where in the second line we specialized to the three solution case. 
For the singlet, the free energy is 
\be
F_{\rm singlet} (T)  = - T \ln (N^2-1)\ .
\ee

\subsection{Conditions for stability  }
  
From \eqn{Lx}, using \eqn{hjk19} and \eqn{raai}, we obtain for the matrix  $\L_{ij}$ the expression
\be 
\label{ladj}
\Lambda_{ij} = { z_i\, \big(\sum_{j=1}^N z_j^{-1}\big) + z_i ^{-1} \Big(\sum_{j=1}^N z_j \Big) \over \r^2-1}\, \delta_{ij} -{z_i z_j^{-1} + z_j z_i^{-1} \over \r^2 -1}    - x_i x_j\ .
\ee
Further using \eqn{jfn22} and \eqn{xira}, we obtain the final expression in terms of $\a_i$
\be
\label{lij}
\Lambda_{ij} = {2\over \rho^2-1} \Big(\rho \cosh \a_i  \delta_{ij}  -
\cosh \a_i \cosh \a_j  - {\rho^2+1\over \rho^2-1} \sinh \a_i \sinh \a_j  \Big)\ .
\ee
This is of the form examined in \cite{PShigher} parametrized in terms of a Euclidean signature metric and two $N$-dimensional vectors.

\subsubsection{Stable solutions}

The matrix $\Lambda_{ij}$ has to satisfy the stability conditions \eqn{Lcond}. 
It turns out that the first of them, $\Lambda \geqslant 0$, is trivially satisfied, so we focus on the second one.
For the generic solution corresponding to a YT of $p_1$ rows of length $\ell_1$ followed by $p_2$ rows of length $\ell_2<\ell_1$,
an exhaustive numerical investigation of the stability condition \eqn{Lcond} reveals that such solutions are generically unstable,
with the following exceptions: 
\be
\label{at}
{\rm A}:\hskip 0 cm  \scalebox{.7}{\begin{ytableau}
\none   &       & 
\end{ytableau}}  \cdots \hskip -.5  cm\scalebox{.7}{  \begin{ytableau}
\none   &     
\end{ytableau}} \qquad   {\rm or}\quad 
\scalebox{.7}{ \begin{ytableau}
\none   &     \boldsymbol{ \bullet}   &     \boldsymbol{ \bullet} 
\end{ytableau}}  \cdots \hskip  -.5 cm \scalebox{.7}{ \begin{ytableau}
\none   &      \boldsymbol{ \bullet}
\end{ytableau}}
\ee
and 
\be
\label{bt}
{\rm B}: \scalebox{.7}{ \begin{ytableau}
\none   &     \boldsymbol{ \bullet}   &     \boldsymbol{ \bullet} 
\end{ytableau}}  \cdots \hskip -.5 cm \scalebox{.7}{ \begin{ytableau}
\none   &      \boldsymbol{ \bullet} & &
\end{ytableau}} \cdots \hskip -.5 cm \scalebox{.7}{  \begin{ytableau}
\none  & 
\end{ytableau}}\ ,
\ee
as well the singlet, which we will denote with a thick dot
\be
\label{st}
{\rm Singlet}: \qq \bullet
\ee
These states can be stable, metastable, or unstable depending on the temperature range.
The two YT in the configuration A are of equal length, and the corresponding solutions are degenerate, having 
the same free energy. Configuration B has an equal number of boxes and antiboxes, so their YT has one row of length $\ell_1$ and 
($N-2$)-rows of length $\ell_2=\ell_1/2$. The two configurations in $A$ are related by conjugation, while B is self-conjugate. 
(Some elementary properties of the conjugate YT adopted to our context are included in the appendix \ref{conjuu}.)

 \no
 Hence, solutions A represent a phase of spontaneously broken conjugation symmetry,
 while B a phase of unbroken conjugation symmetry. The corresponding $SU(N)$ symmetry breaking patterns are
 \be
 \label{fjhja}
 \begin{split}
 &{\rm A:}\quad SU(N) \to SU(N-1) \times U(1) \ ,
 \cr
 &{\rm B:}\quad\, SU(N) \to SU(N-2) \times U(1) \times U(1)\ .
 \end{split}
 \ee
 
\begin{figure}[!ht]
  \begin{center}
    \includegraphics[width=0.99\textwidth]{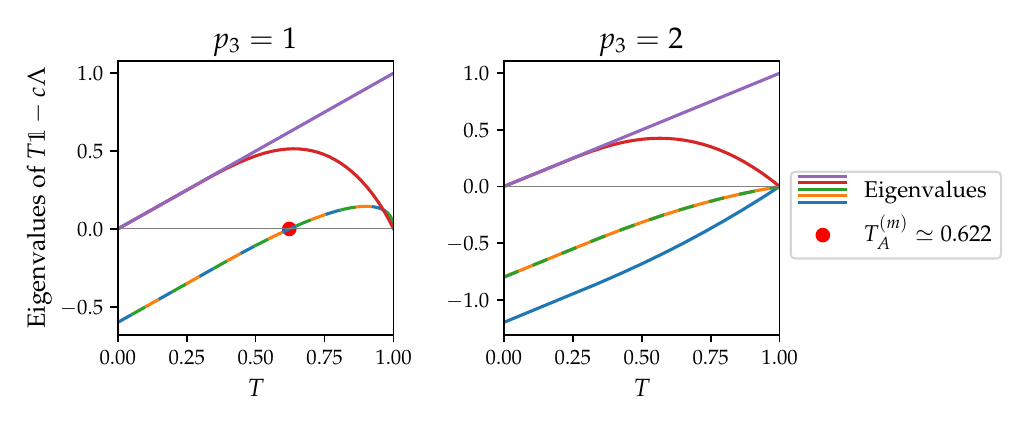}
  \end{center}
  \vskip -1.3 cm
  \caption{\small{Eigenvalues of $T\mathbb{1}-c\Lambda$ for a solution of type A and $N=5$, for $p_3=1$ (left) and $p_3=2$ (right). In the former case, 
  all eigenvalues are positive for temperatures $T_{\rm A}^{\rm (m)} < T < 1$, whereas in the latter one several eigenvalues are negative in the 
  full temperature range in which the solutions exist. Temperatures are in units of $T_0$, and multi-colored lines represent degenerate eigenvalues.}}
  \label{fig:eigenvalues_A_N5}
\end{figure}

\begin{figure}[!ht]
  \begin{center}
    \includegraphics[width=0.99\textwidth]{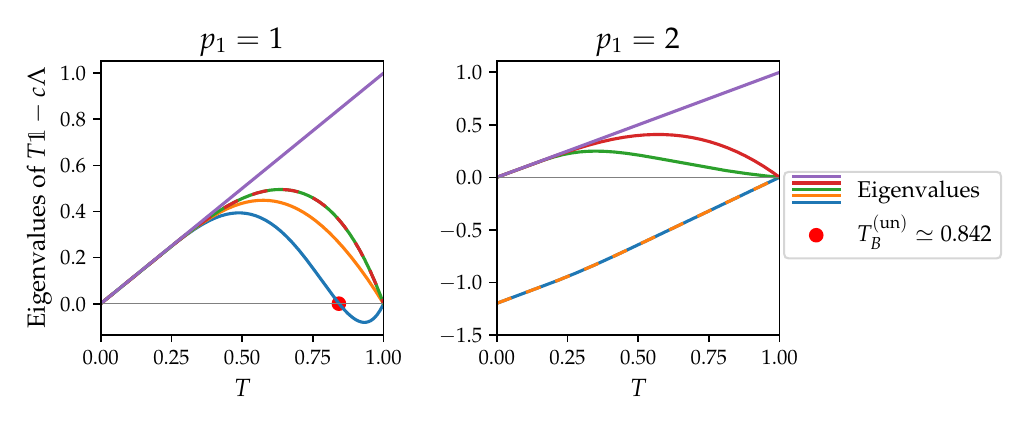}
  \end{center}
  \vskip -1.3 cm
\caption{\small{Eigenvalues of $T\mathbb{1}-c\Lambda$ for a solution of type B and $N=5$, for $p_1=1$ (left) and $p_1=2$ (right). Temperatures are in units of $T_0$. For $p_1=1$, the solution is stable for temperatures below $T_{\rm B}^{\rm (un)}$, whereas for $p_1 = 2$ 
the solution is unstable for all temperatures. Multi-colored lines represent degenerate eigenvalues.}} 
  \label{fig:eigenvalues_B_N5}
\end{figure}

 \no
We stress that there are solutions other than those of type A and type B, but they are unstable.
 For instance, solutions with $p_1=2$ and $p_2=N-4$ are self-conjugate but unstable.
 To illustrate this, we plotted the eigenvalues of the stability matrix $T\, \mathbb{1} - c \L$ in 
 \eqn{Lcond} as a function of the temperature for solutions of type A (fig. \ref{fig:eigenvalues_A_N5}) and type B (fig. \ref{fig:eigenvalues_B_N5}),
 for $p_1=1$ and their extensions for $p_1=2$. We have also examined other non-self conjugate solutions, 
 some of them given in appendix \ref{overkill}, and they turn out to be unstable as well.
 
 \vskip 0 cm
\no
The fact that only configurations A and B with a single row and a single antirow are stable can qualitatively be traced to the fact that, when the atoms carry
the fundamental representation, only one-row YTs are stable \cite{PSferro}. 
 Similarly, if the atoms carry the antifundamental representation, only the conjugate one-antirow YTs  are 
 stable. Since the adjoint can be thought of as a direct product of the fundamental and the antifundamental representations with the singlet removed
 (see section \ref{reducc}),  A and B are solutions that incorporate the aforementioned one-row properties.
 
   \section{Detailed analysis of the stable solutions} \label{detailed}
  
\subsection{Solutions of type A }

\begin{figure} [h]
\begin{center}
\includegraphics[height= 5 cm, angle=0]{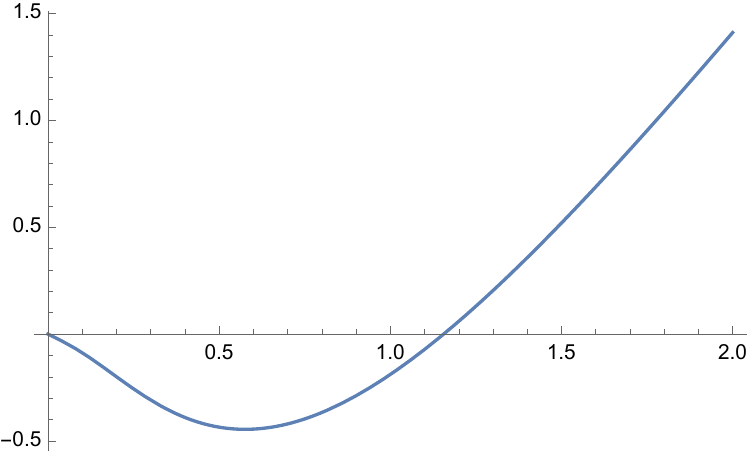}
\end{center}
\vskip -.6 cm
\caption{\small{Plot of the left hand side of \eqn{u13w1} for $T<T_0$.}}  
\label{TlessT0SolA}
\end{figure}

First we examine the case in which the $\a_i$ have only two distinct values. This corresponds to, say, $p_2=0$, with $p_1$ of the $\a_i$
at solution $\a_1>0$ and $p_3=N-p_1$ at the solution $\a_3 <0$. Then the constraint \eqn{u14} gives for $\a_3$
\be
\a_3 =-\ln\Big(f+ \sqrt{ f^2+1}\Big)\ ,\qq  f= {N-p_3\ov p_3} \sinh \a_1
\ee
and $\a_1$ is  determined from the  transcendental equation
\be
\label{u13w1}
 \a_1 - \a_3 - {N T_0(\r)\ov p_3 T} \sinh \a_1= 0 \ ,
\ee
where $T_0(\r)$ is as in \eqn{tor} with 
\be
\r = (N-p_3) e^{\a_1}  + p_3\big(-f+\sqrt{1+f^2}\big)\ .
\ee
The case $p_1 =1$ correspond to a single row, and the case $p_3=1$ to a single antirow, as in states A.
Focusing on the one-antirow stable solution in A with $p_3=1$, the left hand side of \eqn{u13w1} is positive for $\a_1\gg 1$,
while an expansion in $\a_1\ll 1$ yields
\be
\bigg(1-{T_0\ov T}\bigg)\a_1 - 
{(N^2 -1)(N-2)T -(3 N^2 -N+2) T_0 \ov 6  (N+1) T} \a_1^3 + {\cal O}(\a_1)^5 = 0 \ .
\ee
For $T< T_0$ the first term starts out negative, and given the behavior for $\a_1\gg 1$, a solution must exist for a finite value of $\a_1$ 
as depicted in fig. \ref{TlessT0SolA}.

%%%%%%
%%%%%%

\begin{figure} [h]
\begin{center}
\includegraphics[height= 4.5 cm, angle=0]{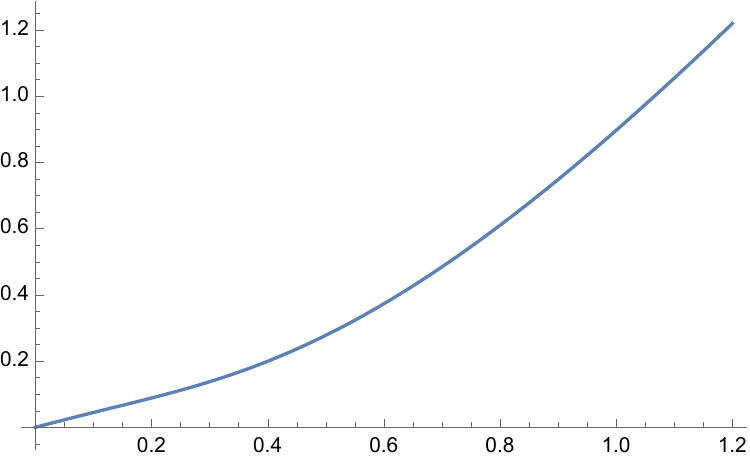}\hskip 1 cm
\includegraphics[height= 4.5 cm, angle=0]{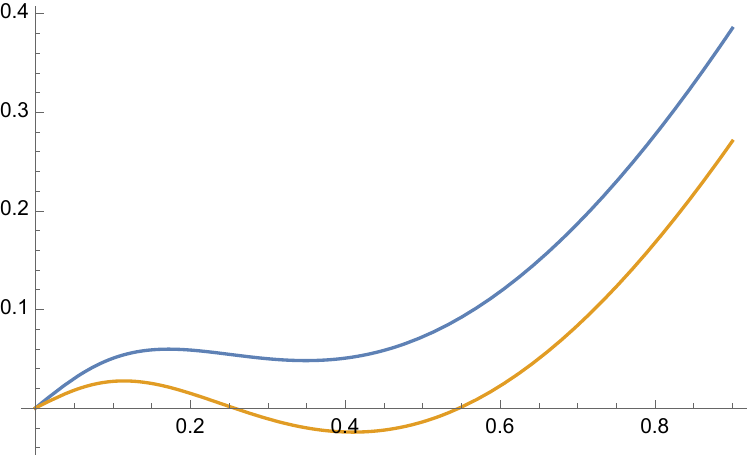}
\end{center}
\vskip -.6 cm
\caption{\small{Left plot: the left hand side of \eqn{u13w1} for $T>T_0$, for $ N=3,4,5$.
Right plot:  the left hand side of \eqn{u13w1} for $N \geqslant 6$  for $T_0  < T^{(c)}_A < T $ (in blue) and for  $T_0 < T<T^{(c)}_A$ (in yellow).}}
\label{TgreaterT0}
\end{figure}

\no
For  $T> T_0$ the first term starts out positive. In order for a non-vanishing solution to exist the coefficient of the cubic term needs to be negative (assuming 
that the first zero occurs for small $\a_1$ which is the case for $T$ slightly larger than $T_0$). Setting $T=T_0$ in this term, we see that it
turns negative for $N>5$. Hence for small enough $T> T_0$ and $N\geqslant 6$, \eqn{u13w1} will have a nonzero solution.
We may compute the critical temperature $T^{(c)}_A$ below which this solution exists by simultaneously setting \eqn{u13wk2} and its
first derivative with respect to $\a_1$ to zero. That gives a system of two transcendental equations 
for $T^{(c)}_A$  and $\a_{1c}$ which has solutions only for  $N\geqslant 6$.
The various cases are depicted in fig. \ref{TgreaterT0}
%%%%%%
%%%%%%

\no
The free energy \eqn{freeF} becomes (for general $p_3$)
\be\label{freee}
F(T)   = T_0\,  {\r^2 (N^2-1) (N-p_3) \ov p_3 (\r^2-1)^2}\sinh^2 \a_1- T\, \ln (\r^2-1)\ .
\ee

%\no The value of $T_c$ cannot be computed analytically but for large $N$ we may show that 
% \be N\gg 1:\qq e^{\a_{1c}}\simeq  \ln^{1/2} (N\ln N) \ ,\qq T_c\simeq {N\ov 2 \ln N}\ . \ee

\subsection{Solutions of type B }

Consider next the case where one of the three solutions of  \eqn{eqq1} is zero, say $\a_2=0$, which enforces $\m=0$.
Then from \eqn{eqq1} and  \eqn{u14} we have that $p_1\a_1+p_3\a_3=0$. This and \eqn{u14} imply 
\be
\sinh {{p_1\ov p_3}\a_1} = {p_1\ov p_3} \sinh \a_1\ ,
\ee
which, for $\a_1 \neq 0$, can only be satisfied if $p_3=p_1$. This implies $\a_3= -\a_1$, and \eqn{eqq1} becomes
\be
\label{u13wk}
 \a_1- {T_0(\r)\ov T} \sinh \a_1 = 0\ ,
\ee
where $T_0(\r)$ is given by \eqn{tor} with 
\be
\label{rrr}
\r =  N +   4 p_1 \sinh^2{\a_1\ov 2}  \ , \qq  T_0(\r) = 2 T_0 {  N +   4 p_1 \sinh^2{\a_1\ov 2}\ov (  N +   4 p_1 \sinh^2{\a_1\ov 2})^2-1} \ .
\ee
We eventually obtain the transcendental equation
\be
\label{u13wk2}
 \a_1- { N^2-1\ov N} {T_0\ov T}{  N +   4 p_1 \sinh^2{\a_1\ov 2}\ov (  N +   4 p_1 \sinh^2{\a_1\ov 2})^2-1} \sinh \a_1 = 0\ .
\ee
Focusing on the stable solution in B with $p_1=1$, the analysis proceeds as in case A.
For $\a_1\gg 1$ the left hand side of \eqn{u13wk2} is positive, while an expansion in $a_1\ll 1$ yields
\be
\bigg(1-{T_0\ov T}\bigg)\a_1 - {T_0\ov T}  {N^3 - 6 N^2 -N-6\ov 6 N (N^2-1)} \a_1^3 + {\cal O}(\a_1)^5 = 0 \ .
\ee
For $T< T_0$ the first term starts out negative, and given the behavior for $a_1\gg 1$, a solution must exist for a finite value of $\a_1$.
For  $T> T_0$ the first term starts out positive. In order for a solution to exist the coefficient of the cubic term needs to be negative (assuming 
that the first zero occurs for small $\a_1$ which is the case for $T$ slightly larger than $T_0$). We see that it
turns negative for $N>6$. Hence, for $N\geqslant 7$ and for small enough $T > T_0$, \eqn{u13wk2} will have a nonzero solution.
We may compute the critical temperature $T^{(c)}_B$ below which this solution exists by simultaneously setting \eqn{u13wk2} and its
first derivative with respect to $\a_1$ to zero. That gives a system of two transcendental equations 
for $T^{(c)}_A$  and $\a_{1c}$ which has solutions only for  $N\geqslant 7$.
The plots of the left hand side of \eqn{u13wk2} are qualitatively
similar to those in fig. \ref{TlessT0SolA} and fig. \ref{TgreaterT0}.
%The reason for this coincidence is that the condition on $N_c$ by solving the above system should result to a single number no matter 
%what the temperature $T$ is, as long as  $T> T_0^{\rm adj}$.

\no
The free energy \eqn{freeF} becomes (for $p_1=1$)
\be
\label{freeF12}
\bb\bb F (T)   = 2 T_0\,  {N^2\hb-\hb 1\ov N}  \Bigg(\bb {\big(N+4 \sinh^2{\a_1\ov 2}\big)   \sinh \a_1  \ov  \big(N+4 \sinh^2{\a_1\ov 2}\big)^2-1)}\bb\Bigg)^2  
\bb- T \, \ln\hb \Big(\hb\big(N+4 \sinh^2{\a_1\ov 2}\big)^2-1\hb\Big)\, .
\ee

%%%%%%%%%%
\subsection{Phases and transitions }

At low temperatures, the stable configuration is B for all values of $N$. Similarly, for high temperatures the singlet is the 
only solution and it is stable. For intermediate temperatures and $N\geqslant 4$, the conjugation symmetry breaking phase A also appears,
and configurations can become stable, metastable, or cease to exist. For $N=3$, the situation is simple and presented in table \ref{table:N=3}.
For $N\geqslant 4$, there are subtle differences depending on the values of $N$, and unexpectedly numerous cases appear, which we present
below in detail. Focusing on the stable phases only, the presentation simplifies and is given in table  \ref{table:N=45small} for $N=4,5$ and in table \ref{table:6...} for $N\geqslant 6$.

\subsubsection{The $SU(2)$ case} 
\label{adjsu2}

In \cite{PSspins} the decomposition of $n$ spin-$s$ irreps of $SU(2)$ was presented and a ferromagnetic model similar to the
one we consider here was studied in the mean field approximation. 
Since $s=1$ corresponds to the adjoint $SU(2)$ irrep, the $s=1$ results should match those of the present paper.
Indeed, letting $s=1$ in eq. (4.20) of  \cite{PSspins} and also rescaling $c\to 2 c$ in order to match conventions, we obtain
\be\label{s1}
{2 \sinh \tau \ov 2\cosh \tau +1} = {\tau T\ov 2 c}\ .
\ee 
A non-vanishing solution for $\tau$ implies a magnetized configuration.  
 \eqn{s1} becomes \eqn{u13wk2} upon setting $N=2$, $c={3T_0/4}$, $p_1=1$ and defining $\tau = 2 \a_1$. The corresponding one-row YT derived in
 \cite{PSspins} also matches the one derived in the present paper.

\subsubsection{The $SU(3)$ case} 

For $SU(3)$ we have a simple picture, depicted in table \ref{table:N=3}. In this case, configuration A is never stable for $T<T_0$.
\begin{table}[!ht]
\begin{center}
\begin{tabular}{|c||c|c|} 
\hline
  irrep & $T<T_0$ & $T_0<T$ \\ 
\hline\hline
$\bullet$  &  {\rm unstable}    & {\rm stable} \\ 
\hline
B & {\rm stable}    & $\times $\\ 
\hline
A & {\rm unstable}    & $\times $\\ 
\hline
\end{tabular}
\end{center}
\vskip -0.4cm
\caption{\small{
Phases in various temperature ranges for $N =3$ and their stability characterization.
}}
\label{table:N=3}
\end{table}
We note that in this case the YT that corresponds to configuration B is simply 
\be
{\rm B}\ {\rm for}\ SU(3):\scalebox{.7}{ \begin{ytableau}
\none   &  &   \\
\none   & &
\end{ytableau}}  \cdots \hskip -.5 cm \scalebox{.7}{ \begin{ytableau}
\none   &      & & \\
\none & 
\end{ytableau}} \cdots \hskip -.5 cm \scalebox{.7}{  \begin{ytableau}
\none  & 
\end{ytableau}}\ .
\ee

\subsubsection{The $SU(4)$ and $SU(5)$ cases} 

The situation is more complicated, with several critical temperatures listed from low to high: a temperature
$T^{\rm (m)}_{\rm A}$ at which the configuration A becomes from unstable metastable; a higher temperate $T_{\rm BA}$ at which configurations B and A
exchange their metastability; and an even higher temperature $T^{\rm (un)}_{\rm B}$ at which configuration B becomes unstable. All these 
temperatures are smaller than $T_0$, above which only the singlet exists and is stable. 
We have tabulated the results in table  \ref{table:N=45}. 

\begin{table}[!ht]
\begin{center}
\begin{tabular}{|c||c|c|c|} 
\hline
Temperature/irrep  & 
B &
A &
$\bullet$ \\
\hline\hline
$T < T^{\rm (m)}_{\rm A}$ & {\rm stable} & {\rm unstable} & {\rm unstable} \\
\hline
$T^{\rm (m)}_{\rm A} < T < T_{\rm BA}$ & {\rm stable} & {\rm metastable} & {\rm unstable} \\
\hline
$T_{\rm BA} < T < T^{\rm (un)}_{\rm B}$ & {\rm metastable} & {\rm stable} & {\rm unstable} \\
\hline
$T^{\rm (un)}_{\rm B} < T < T_0$ & {\rm unstable} & {\rm stable} & {\rm unstable} \\
\hline
$T_0 < T$ & $\times$ & $\times$ & {\rm stable} \\
\hline
\end{tabular}
\end{center}
\vskip -0.4cm
\caption{\small{
Phases in various temperature ranges for $ N =4,5$ and their stability.
}}
\label{table:N=45}
\end{table}
\no
Table \ref{table:N=45small} depicts only the stable configurations in each temperature range.
\begin{table}[!ht]
\small
\begin{center}
\begin{tabular}{|c||c|c|c|}
\hline
Temperature & $T<T_{\rm BA}$ & $T_{\rm BA} < T < T_0$ & $T_0 < T$ \\
\hline
irrep
& B & A & $\bullet$   \\
\hline
\end{tabular}
\end{center}
\vskip -0.4cm
\caption{\small
Stable phases in the various  temperature ranges for $N=4,5$.
}
\label{table:N=45small}
\end{table}

\no
For all cases up to $N=5$ there is a phase transition at $T= T_0$. We may compute the order of the phase transition by 
examining the free energy of the solutions at this transition temperature.  For the cases with $N=4,5$ we have a transition from the solution A to the singlet
so that we may use for the free energy that in \eqn{freee} for $p_3=1$. For $N=3$ we may use the free energy in \eqn{freeF12} for the B solution. 
For $N=2$ either \eqn{freee} or  \eqn{freeF12}  since the A and B solutions coincide. 
%\be
%\label{freeF1}
%F(T)   = T_0\,  {\r^2 \ov (\r^2-1)^2}  (N-1)(N^2-1)  \sinh^2 \a_1 - T\, \ln (\r^2-1)\ .
%\ee
For $T \lesssim  T_0$ we have the expansions
\be
\begin{split} 
& \a_1  = c_1 \sqrt{T_0-T}\,  \Big(1+ c_2 (T_0-T) + \dots \Big)\ ,
\\
& N=2 : \quad c_1= \sqrt{3\ov 2}\ ,\quad c_2 = {13\ov 20}\ , 
\\
& N=3 : \quad  c_1= 2\ ,\quad c_2 = {11\ov 15}\ ,
\\
& N=4 : \quad c_1= \ha \sqrt{15\ov 2} \ ,\quad c_2 = {9\ov 8}\ ,
\end{split}
\ee
where we have used the either the equation for the A solution or the B solution (for $N=3$).
This gives the expansion of the free energy for $T \lesssim  T_0$ which compactly takes the form
\be
F(T)  = - T \ln(N^2-1) + {3\ov N-5} (T-  T_0)^2+ \dots\ ,\quad  N=2,3,4\ .
\ee
For $T>T_0$, the free energy of the singlet is $-T\ln (N^2-1)$.
Hence, $T_0$ marks a second order phase transition.

\no
For $N=5$ things are a bit different. We find that 
\be
 \a_1  = {\sqrt{6}\ov 5^{3/4}} (T_0-T)^{1/4}\,   \Big(1+ {461\ov 140 \sqrt{5}} \big(T_0-T\big)^{1/2} + \dots \Big)\ .
 \ee
Then the expansion of the free energy for $T \lesssim  T_0$ is
\be
\label{jfhj22}
F (T)  = - T \ln 24 - {2 \sqrt{5}\ov 3} \big(T_0-T\big)^{3/2}+ \dots\ ,\quad  N=5\ .
\ee
Hence, the transition can be characterized as a second order one, albeit the second derivative becomes infinite.

\no
The metastability transition at $T_{\rm BA}$ is first order, as is generically the case.

\subsubsection{The $SU(6)$ case}  
In this case, configuration A  exists for temperatures above $T_0$ up to a temperature $T_{\rm A}^{(c)}$, and
there is a  temperature $T_{\rm AS}$ such that $T_0 < T_{\rm AS}< T_{\rm A}^{(c)} $ at which A and the singlet exchange metastabilty roles.
The results are  presented in table  \ref{table:N=6}. 

\begin{table}[ht!]
\begin{center}
\begin{tabular}{|c||c|c|c|} 
\hline
Temperature/irrep   & B & A & $\bullet$ \\ 
\hline\hline
$T < T^{\rm (m)}_{\rm A} $ & {\rm stable} & {\rm unstable} & {\rm unstable} \\
\hline
$T^{\rm (m)}_{\rm A} < T < T_{\rm BA}$ & {\rm stable} & {\rm metastable} & {\rm unstable} \\
\hline
$T_{\rm BA}< T < T^{\rm (un)}_{\rm B}$ & {\rm metastable} & {\rm stable} & {\rm unstable} \\
\hline
$T^{\rm (un)}_{\rm B} < T < T_0$ & {\rm unstable} & {\rm stable} & {\rm unstable} \\
\hline
$T_0<T<T_{\rm AS}$ & $\times$ & {\rm stable} & {\rm metastable} \\ 
\hline
$T_{\rm AS} <T<T_{\rm A}^{(c)}  $ & $\times$ & {\rm metastable} & {\rm stable} \\ 
\hline
$T_{\rm A}^{(c)}  <T$ & $\times$ & $\times$ & {\rm stable} \\ 
\hline
\end{tabular}
\end{center}
\vskip -0.3cm
\caption{\small{
Phases in various temperature ranges for $N = 6$ and their stability.
}}
\label{table:N=6}
\end{table}

\no
The phase transition  between the B and the A solutions at $T_{\rm BA}$ is first order similar to the cases with $N\leqslant 5$.
However, unlike these cases, for $N=6$ the phase transition between the A solution and
the singlet is first order, instead of second, and is happening at $T_{\rm AS}>T_0$. This feature persists for all higher values of $N$.

\no
We may construct the smaller table \ref{table:6...} with only the stable configurations in each temperature range. In fact this covers all cases 
with $N\geqslant 6$ and will not be repeated subsequently.
\begin{table}[!ht]
\small
\begin{center}
\begin{tabular}{|c||c|c|c|c|}
\hline
Temperature & $T < T_{\rm BA} $ & $T_{\rm BA}  < T <T_{\rm AS}$ & $T_{\rm AS}< T$ \\
\hline
irrep & B & A & $\bullet $  \\
\hline
\end{tabular}
\end{center}
\vskip -0.3cm
\caption{\small
Stable phases in the various temperature ranges for $N\geqslant 6 $. }
\label{table:6...}
\end{table}

\no
This is similar to table \ref{table:N=45small}, the only difference being that the singlet takes over at the temperature $T_{\rm AS}> T_0$. So the result of  tables \ref{table:6...} and  \ref{table:N=45small} is that for high temperatures the $SU(N)$ symmetry as well as the conjugation symmetry 
are intact, for intermediate temperatures both are broken, whereas for low enough temperatures conjugation is restored.  The breaking patterns are given in 
\eqn{fjhja}.

\subsubsection{The $SU(7)$ case} 
In this case the configuration B also exists above $T_0$ up to a temperature $T_{\rm B}^{(c)}$, with 
 $T_0 < T_{\rm B}^{(c)}< T_{\rm AS}$. 
The results are  presented in table  \ref{table:N=7}. 

\begin{table}[!ht]
\begin{center}
\begin{tabular}{|c||c|c|c|} 
\hline
Temperature/irrep & B & A & $\bullet$ \\ 
\hline\hline
$T < T^{\rm (m)}_{\rm A} $ & {\rm stable} & {\rm unstable} & {\rm unstable} \\
\hline
$T^{\rm (m)}_{\rm A}  < T < T_{\rm BA} $ & {\rm stable} & {\rm metastable} & {\rm unstable} \\
\hline
$T_{\rm BA} < T < T^{\rm (un)}_{\rm B} $ & {\rm metastable} & {\rm stable} & {\rm unstable} \\
\hline
$T^{\rm (un)}_{\rm B}  < T < T_0$ & {\rm unstable} & {\rm stable} & {\rm unstable} \\
\hline
$T_0<T<T_{\rm B}^{(c)}$ & {\rm unstable} & {\rm stable} & {\rm metastable} \\ 
\hline
$T_{\rm B}^{(c)} <T<T_{\rm AS}$ & $\times$ & {\rm stable} & {\rm metastable} \\ 
\hline
$T_{\rm AS}<T< T_{\rm A}^{(c)} $ & $\times$ & {\rm metastable} & {\rm stable} \\ 
\hline
$T_{\rm A}^{(c)} <T$ & $\times$ & $\times$ & {\rm stable} \\ 
\hline
\end{tabular}
\end{center}
\vskip -0.4cm
\caption{
Phases in various temperature ranges for $N = 7$ and their stability.
}
\label{table:N=7}
\end{table}
\no
Note the ordering of temperatures $ T_{\rm BA} < T^{\rm (un)}_{\rm B}< T_0$. This ordering will change with higher values of $N$.

\no
In general, for $N\geqslant 6$ the eigenvalues of the  stability matrix $T\, \mathbb{1} - c \L$ as a function of the temperature have similar plots to
those in fig. \ref{fig:eigenvalues_A_N5} and fig. \ref{fig:eigenvalues_B_N5}, with the difference that they extend to temperatures higher than $T_0$,
and will not be presented here.

\subsubsection{The $SU(8)$ case} 

As noted, the ordering of the various critical temperatures changes as $N$ increases, until it stabilizes for $N\geqslant 13$.
In  the case of $N=8$ the ordering of temperatures changes to $ T_{\rm BA} < T_0< T^{\rm (un)}_{\rm B}$. The results are  presented in table  \ref{table:N=8}. 

\begin{table}[!ht]
\begin{center}
\begin{tabular}{|c||c|c|c|} 
\hline
Temperature/irrep  & B & A & $\bullet$ \\ 
\hline\hline
$T< T^{\rm (m)}_{\rm A}  $ & {\rm stable} & {\rm unstable} & {\rm unstable} \\ 
\hline
$T^{\rm (m)}_{\rm A}  <T<T_{\rm BA}$ & {\rm stable} & {\rm metastable} & {\rm unstable} \\ 
\hline
$T_{\rm  BA} <T<T_0$ & {\rm metastable} & {\rm stable} & {\rm unstable} \\ 
\hline
$T_0<T< T^{\rm (un)}_{\rm B} $ & {\rm metastable} & {\rm stable} & {\rm metastable} \\ 
\hline
$T^{\rm (un)}_{\rm B}  < T < T_{\rm B}^{(c)} $ & {\rm unstable} & {\rm stable} & {\rm metastable} \\ 
\hline
$T_{\rm B}^{(c)}  < T < T_{\rm AS} $ & $\times$ & {\rm stable} & {\rm metastable} \\ 
\hline
$T_{\rm  AS}  < T < T_{\rm A}^{(c)}$ & $\times$ & {\rm metastable} & {\rm stable} \\ 
\hline
$T_{\rm A}^{(c)} < T$ & $\times$ & $\times$ & {\rm stable} \\ 
\hline
\end{tabular}
\end{center}
\vskip -0.4cm
\caption{
Phases in various temperature ranges for $ N =8$ and their stability.
}
\label{table:N=8}
\end{table}

\subsubsection{The cases $SU(N)$ with $N=9,10,11$}  

In  this case the ordering of temperatures changes to $T_0 < T_{\rm BA}  < T^{\rm (un)}_{\rm B}$. 
The results are  presented in table  \ref{table:N=91011}. 

\begin{table}[!ht]
\begin{center}
\begin{tabular}{|c||c|c|c|} 
\hline
Temperature/irrep  & B & A & $\bullet$ \\ 
\hline\hline
$T<T^{\rm (m)}_{\rm A} $ & {\rm stable} & {\rm unstable} & {\rm unstable} \\ 
\hline
$T^{\rm (m)}_{\rm A} <T<T_0$ & {\rm stable} & {\rm metastable} & {\rm unstable} \\ 
\hline
$T_0<T<T_{\rm BA} $ & {\rm stable} & {\rm metastable} & {\rm metastable} \\ 
\hline
$T_{\rm BA} <T<T^{\rm (un)}_{\rm B} $ & {\rm metastable} & {\rm stable} & {\rm metastable} \\ 
\hline
$T^{\rm (un)}_{\rm B}  < T < T_{\rm B}^{(c)}  $ & {\rm unstable} & {\rm stable} & {\rm metastable} \\ 
\hline
$T_{\rm B}^{(c)}   < T < T_{\rm AS} $ & $\times$ & {\rm stable} & {\rm metastable} \\ 
\hline
$T_{\rm AS}  < T <  T_{\rm A}^{(c)} $ & $\times$ & {\rm metastable} & {\rm stable} \\ 
\hline
$ T_{\rm A}^{(c)} < T$ & $\times$ & $\times$ & {\rm stable} \\ 
\hline
\end{tabular}
\end{center}
\vskip -0.3cm
\caption{
Phases in various temperature ranges for $N=9,10,11$ and their stability.
}
\label{table:N=91011}
\end{table}
\no
Note the ordering of temperatures $ T^{\rm (c)}_{\rm B}< T_{\rm AS}$ which will change subsequently.

\subsubsection{The $SU(12)$ case}  

In  this case the ordering changes to  $T_{\rm AS} < T^{\rm (c)}_{\rm B}$. The results are presented in table  \ref{table:N=12}. 

\begin{table}[!ht]
\begin{center}
\begin{tabular}{|c||c|c|c|} 
\hline
Temperature/irrep  & B & A & $\bullet$ \\ 
\hline\hline
$T< T^{\rm (m)}_{\rm A} $ & {\rm stable} & {\rm unstable} & {\rm unstable} \\ 
\hline
$T^{\rm (m)}_{\rm A} <T<T_0$ & {\rm stable} & {\rm metastable} & {\rm unstable} \\ 
\hline
$T_0<T<T_{\rm BA}$ & {\rm stable} & {\rm metastable} & {\rm metastable} \\ 
\hline
$T_{\rm BA} < T < T^{\rm (un)}_{\rm B} $ & {\rm metastable} & {\rm stable} & {\rm metastable} \\ 
\hline
$ T^{\rm (un)}_{\rm B}  < T < T_{\rm AS} $ & {\rm unstable} & {\rm stable} & {\rm metastable} \\ 
\hline
$T_{\rm AS}  < T <  T_{\rm B}^{(c)} $ & {\rm unstable} & {\rm metastable} & {\rm stable} \\ 
\hline
$T_{\rm B}^{(c)}  < T < T_{\rm A}^{(c)}  $ & $\times$ & {\rm metastable} & {\rm stable} \\ 
\hline
$T_{\rm A}^{(c)}   < T$ & $\times$ & $\times$ & {\rm stable} \\ 
\hline
\end{tabular}
\end{center}
\vskip -0.3cm
\caption{
Phases in various temperature ranges for $N = 12$ and their stability.
}
\label{table:N=12}
\end{table}

%%%%%
\subsubsection{The $SU(N)$ case with $N\geqslant 13 $}  

In  this case the ordering changes to  $T_{\rm AS} < T^{\rm (un)}_{\rm B}$. The results are presented in table  \ref{table:N>=13}. 

\begin{table}[!ht]
\begin{center}
\begin{tabular}{|c||c|c|c|}
\hline
Temperature/irrep  & B & A & $\bullet$ \\
\hline\hline
$T<T_{\rm A}^{\rm (m)}$ & {\rm stable} & {\rm unstable} & {\rm unstable} \\
\hline
$T_{\rm A}^{\rm (m)}<T<T_0$ & {\rm stable} & {\rm metastable} & {\rm unstable} \\
\hline
$T_0<T<T_{\rm BA}$ & {\rm stable} & {\rm metastable} & {\rm metastable} \\
\hline
$T_{\rm BA}<T<T_{\rm AS}$ & {\rm metastable} & {\rm stable} & {\rm metastable} \\
\hline
$T_{\rm AS} < T < T_{\rm B}^{\rm (un)}$ & {\rm metastable} & {\rm metastable} & {\rm stable} \\
\hline
$T_{\rm B}^{\rm(un)} < T < T_{\rm B}^{(c)}$ & {\rm unstable} & {\rm metastable} & {\rm stable} \\
\hline
$T_{\rm B}^{(c)} < T < T_{\rm A}^{(c)}$ & $\times$ & {\rm metastable} & {\rm stable} \\
\hline
$T_{\rm A}^{(c)} < T$ & $\times$ & $\times$ & {\rm stable} \\
\hline
\end{tabular}
\end{center}
\vskip -0.3cm
\caption{
Phases in various temperature ranges for $N \geqslant 13$ and their stability.
}
\label{table:N>=13}
\end{table}

\subsubsection{$SU(N)$ in the  $N\gg 1 $ limit}  

The limiting behavior of the various temperatures for $N\gg 1$ is
\be
\label{eq:N>>1_temps}
\begin{split}
& T^{\rm (m)}_{\rm A} \simeq  {T_0\ov 2} \ , 
\\
&T_{\rm BA} \simeq T_{\rm AS} \simeq  T_0 {N\ov 4 \ln N}  \ , 
\\
& T^{\rm (un)}_{\rm B}\simeq  T_{\rm A}^{(c)}  \simeq T_{\rm B}^{(c)}  \simeq T_0 {N\ov 2 \ln N}\ .
\end{split} 
\ee
These have been found using a combination of analytical and numerical methods. 
We note that even though $N\gg 1$ we are still in the regime $N\ll n$.
In this limit, we have $T_{\rm B}^{\rm (un)} \simeq T_{\rm B}^{(c)} \simeq T_{\rm A}^{(c)}$, and $T_{\rm BA} \simeq T_{\rm AS}$. 
The last equality implies that the range in which phase A is globally stable shrinks, and B survives as the only stable phase other than the
singlet. The single temperature $T_{\rm BA} \simeq T_{\rm AS}$ marks the transition between B and the singlet,
hence we will denote it by $T_{\rm BS}$. The transition table simplifies as below.
\begin{table}[!ht]
\begin{center}
\begin{tabular}{|c||c|c|c|} 
\hline
Temperature/irrep & B & A & $\bullet$ \\ 
\hline\hline
$T<T_{\rm A}^{\rm (m)}$ & {\rm stable} & {\rm unstable} & {\rm unstable} \\ 
\hline
$T_{\rm A}^{\rm (m)}<T<T_0$ & {\rm stable} & {\rm metastable} & {\rm unstable} \\ 
\hline
$T_0<T<T_{\rm BS}$ & {\rm stable} & {\rm metastable} & {\rm metastable} \\ 
\hline
$T_{\rm BS}<T<T_{\rm A}^{(c)}$ & {\rm metastable} & {\rm metastable} & {\rm stable} \\ 
\hline
$T_{\rm A}^{(c)} < T$ & $\times$ & $\times$ & {\rm stable} \\
\hline
\end{tabular}
\end{center}
\vskip -0.3cm
\caption{
 Phases in various temperature ranges for $N \gg 1$ and their stability.}
\label{table:10}
\end{table}
\no
A smaller table \ref{table:N>>1short} is the analog of table \ref{table:6...} for $N \gg 1$.
\begin{table}[!ht]
\small
\begin{center}
\begin{tabular}{|c||c|c|}
\hline
Temperature & $T < T_{\rm BS} $ & $T_{\rm BS} < T$ \\
\hline
irrep & B & $\bullet$ \\
\hline
\end{tabular}
\end{center}
\vskip -0.3cm
\caption{\small
Stable phases in the various temperature ranges for $N\gg 1$.
}
\label{table:N>>1short}
\end{table}

\no
The evolution of the various critical temperatures $T_{\rm A}^{\rm(m)}, T_{\rm BA}$, etc. as functions of $N$ is plotted in fig. \ref{fig:Tplot} below.
As noted, for $N\gg 1$ there is no intermediate temperature regime in which the solution A is stable.
\begin{figure}[!ht]
	\begin{center}
		\includegraphics[width=0.7\textwidth]{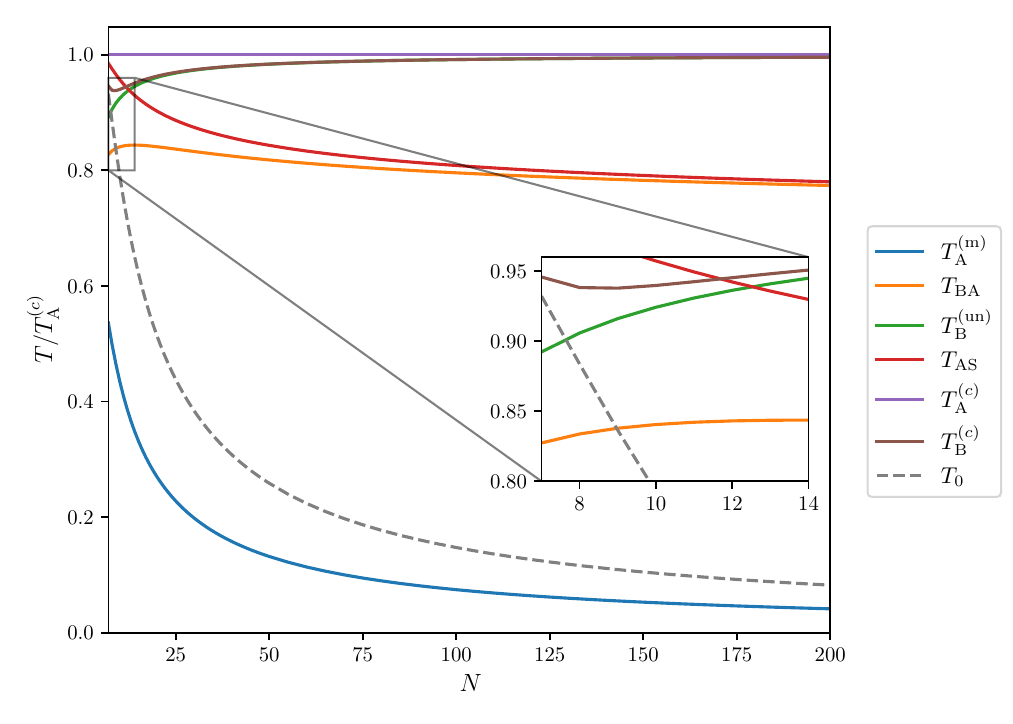}
	\end{center}
	\vskip -.9 cm 
	\caption{\small{Evolution of the temperatures $T_{\rm A}^{\rm (m)}, T_{\rm BA}, T_{\rm B}^{\rm (un)}, T_{\rm AS}, T_{\rm A}^{(c)}$ and $T_{\rm B}^{(c)}$ with the group rank $N$. All temperatures are normalized by $T_{\rm A}^{(c)}$. The inset serves to show the crossings between the different temperatures 
	that occur for $7\leqslant N\leqslant 13$.
According to \eqn{eq:N>>1_temps} $T_{\rm BA}$ and $T_{\rm AS}$ eventually converge to $T_{\rm A}^{\rm (c)}/2$. 
This becomes visible at extremely large $N$, due to the slow convergence of 
$T_{\rm B}^{\rm (un)}$, $T_{\rm A}^{(c)}$ and $T_{\rm B}^{(c)}$ to their large $N$ limit,
and is not depicted in the figure.}} 
	\label{fig:Tplot}
\end{figure}

%%
% I define a command to quickly call the YT
\newcommand{\reducibleYT}{%
  \hskip -.1cm
  \tiny{%
    \begin{ytableau}
      \none & \boldsymbol{\bullet}
    \end{ytableau}
  }%
  \otimes
  \hskip -.3cm
  \tiny{%
    \begin{ytableau}
      \none &
    \end{ytableau}
  }%
}
\subsection{The reducible fundamental-antifundamental representation}
\label{reducc}

To monitor the effects that relatively minor variations of the model have on the results,
we consider atoms in the reducible tensor product representation
 $\reducibleYT = \hskip -.26 cm  \tiny{%
    \begin{ytableau}
      \none & \boldsymbol{\bullet} &
    \end{ytableau}
  } %
  \oplus \mathbb{1}$
which differs form the adjoint by the presence of an additional singlet state per atom. We will see that this affects our results only minimally.

 \no
 Application of the general formalism to the above representation is straightforward. We note that, in the reducible case, the self-interaction terms
 $\sum_{a=1}^{N^2-1} (j_i^a)^2$ that must be eliminated from the mean-field Hamiltonian are not constant Casimirs any more, but their contribution is
 subleading in $n$ and can be neglected.
 The analysis results to equilibrium equations and constraints identical to the ones for the adjoint, upon
 substituting $\r^2 \pm 1 \to \r^2$ and $N^2 -1 \to N^2$ in the expressions.
 This modification leads to a similar phase structure as for the adjoint representation, with some quantitative differences.
 The values of the critical temperatures $T_{\rm A}^{({\rm m})}, T_{\rm B}^{{\rm un})}, T_{\rm BA}, T_{\rm AS}, T_{\rm A}^{(c)}$ and $T_{\rm B}^{(c)}$
 change slightly, but the tables \ref{table:N=3} to \ref{table:N>>1short} remain almost intact. 
 Specifically, tables \ref{table:N=45} and \ref{table:N=45small} hold for $N=4$ only, table \ref{table:N=6} holds for $N=5$ and $6$, table \ref{table:6...}
 holds for $N \geqslant 5$, and all other tables remain unmodified.

\no
Phase transitions are also generally of the same order as in the adjoint case, with the exception of the transition of the A state to the
singlet for $N=5$, which is no longer of second order at $T=T_0$, but becomes first order at $T_{\rm AS} > T_0$.
Further, the nonanalytic behavior in \eqn{jfhj22} of the adjoint case for $N=5$ is now absent.

\no
The case of $SU(2)$ is particularly simple, since its fundamental and antifundamental irreps are equivalent, and thus $n$ reducible representations
$\bb\bb\reducibleYT$ are equivalent to $2n$ fundamentals. Equivalently, it corresponds to $\lambda = 2\lambda_\text{fund}$, and from \eqn{fbx} we deduce
\be
F(c,\bx) = 2\, F_\text{fund} (2c,\bx/2) \ .
\label{ff2n}\ee
Indeed, the equilibrium equation for $x_1 = -x_2 \equiv x$ and the corresponding free energy coincide with the ones in \cite{PSferro} up to the
factors of 2 in \eqn{ff2n}.

\section{Conclusions}\label{conclusions}

Ferromagnets with atoms in the $SU(N)$ adjoint representation prove to be interesting and intricate
physical systems, adding to the already rich array of phenomena of ferromagnets in the fundamental,
symmetric, and antisymmetric representations. The new element is the conjugation invariance of the free energy,
which adds a discrete symmetry to the global $SU(N)$ symmetry of the system. Our results reveal that this symmetry
can also be spontaneously broken, although in a more restricted temperature range that the magnetization
$SU(N)$ symmetry. We expect the spontaneous breaking of conjugation symmetry to be a generic feature of ferromagnets
with atoms in an arbitrary self-conjugate representation.

\no
It is interesting that the phase structure we have unveiled depends to a good extent on the rank $N$ of the group, as the large number of phase transition
patterns indicates.
This dependence appears at nonvanishing temperature and does not seem to have a clear group theoretic origin.

\no
The phase structure of the adjoint ferromagnet shares many of the features of the ferromagnet with atoms in the antisymmetric representation
studied in \cite{PShigher}, which include the appearance of more than two phases, and regimes
in which they coexist as stable and metastable states. The presence of metastable states is physically significant, since their thermally driven
transition to a fully stable state takes an exponentially long time. Thus, metastable states are practically stable if left unperturbed, and only
external perturbations (impurities, shaking the system etc.) can catalyze their decay.

\no
The symmetry breaking patterns of the adjoint ferromagnet validate the general conjectures put forth in our previous work
\cite{PShigher}. One of them was that, if the atom irrep $\chi$ has $r$ rows, the ferromagnet will have
up to $r+1$ distinct phases corresponding to irreps with $0,1,\dots,r$ rows ($0$ rows corresponding to the singlet). 
The adjoint can be thought of as a two-row state, with one row and one antirow, and indeed its stable phases are the singlet,
one row, one antirow, and the (self-conjugate) one row plus one antirow irreps, in agreement with the conjecture.
Further, the critical temperature $T_0^{(\chi)}$ below which the singlet becomes unstable has the general conjectured form
\eqn{cggrp} in terms of group theoretical quantities.
Proving the above conjectures for a general atom representation is an interesting open question.

\no
Physical applications of our adjoint ferromagnet can vary, 
but they fall within the general category of couplings that are not fully invariant under atom state rotations (interactions
are not of purely exchange type). Specific situations where the adjoint is relevant will involve systems with conjugation invariance,
and their experimental realization is of substantial physical interest.

\no
The work in this paper extends the results of \cite{PShigher} in one of several possible directions, namely to a
different higher atom irrep. Coupling the atoms to magnetic fields, which can be studied with the existing formalism,
would be useful for probing the response of the states to external fields.
Other generalizations suggested in \cite{PSferro,PShigher}, e.g., including three- and higher-atom interactions
for the adjoint, as was done for the fundamental in \cite{PScubic}, or inclusion of interaction terms along the Cartan generators
breaking the $SU(N)$ invariance, remain open to investigation. The study of the adjoint
model in the double scaling large-$N$ limit $N \sim \sqrt n$ studied in
\cite{PSlargeN,PSferroN} for fundamental atoms is also interesting. We expect that the large-$N$ limiting 
behavior in the scaling $N \sim \sqrt{n}$ will lead to transitions into "disjoint" phases at large temperature.
Further, the relevance of our results to nonabelian topological phases in one dimension
\cite{TLC,RQ,CFLT,RPAR}
and the persistence of such phases in higher dimensions are interesting questions for future investigation.

\no
In a different direction, our results for $SU(N)$ ferromagnets bear a tantalizing resemblance
to those of $N$-state Potts spin models on a complete graph, recently studied in the context of models for collective communication
and competition in social systems \cite{Nechaev}. The relevance of our adjoint
and other higher irrep ferromagnets to social systems is a fascinating topic that merits exploration.

\no
Finally, extending our ferromagnetic model to other classical groups, supersymmetric groups with fermionic
dimensions, or "quantum" $q$-groups are interesting possibilities and offer avenues
for further research.

%%%%%

\subsection*{Acknowledgements }

A.P. would like to thank Sergei Nechaev for interesting discussions. A.P.'s research was supported by the National Science Foundation 
under grants NSF-PHY-2112729 and NFS-PHY-2112479, and by PSC-CUNY grants 67100-00 55 and 6D136-00 05.
%K.S. would like to thank the Department of Theoretical Physics at CERN for financial support and hospitality during the late stages of this research. 

%%%%\newpage

\appendix 

\section{On conjugate representations} 
\label{conjuu}

Elaborating on the notion of conjugation, consider a YT with $p_i$ rows of length $\ell_i$, $i=1,2,\dots, N-1$.
The conjugate YT is obtained by subtracting the original YT from the $\ell_1 \times N$ rectangle framing it and rotating the
remainder by 180$^o$ (this can be equivalently thought of as turning all rows into antirows).
 As an example, the conjugate of the YT with $p_1$ rows of length $\ell_1$ and $p_2$ rows of length $\ell_2<\ell_1$ (equivalently,
 $p_1$ rows of length $\ell_1-\ell_2$ and $N-p_1-p_2$ antirows of length $\ell_2$)
 has $N-p_1-p_2$ rows of length $\ell_1$ and $p_2$ rows of length $\ell_1-\ell_2$.  
 If $N-p_1-p_2=p_1$ (that is, $2 p_1 + p_2=N$) and $\ell_1-\ell_2=\ell_2$ (that is, $2 \ell_2=\ell_1$),
 then the YT maps to itself under conjugation and is self-conjugate. This is  possible only if $\ell_1$ is even. However, in our case this is 
 immaterial since row lengths scale with the number of atoms $n\gg 1$. Note that the number of 
 boxes in a self-dual YT of the above type is $N\ell_1/2$ and hence independent of $p_1$.

%%%

\section{More generic, unstable solutions}\label{overkill}

Besides the solutions A and B, one may wonder if there are other stable solutions. 
It turns out that there are none. Any modification of these solutions results to instability for all temperatures.

\begin{figure}[!ht]
{}\vskip 1cm
 \begin{center}
   \hskip -2 cm \includegraphics[width=1.1\textwidth]{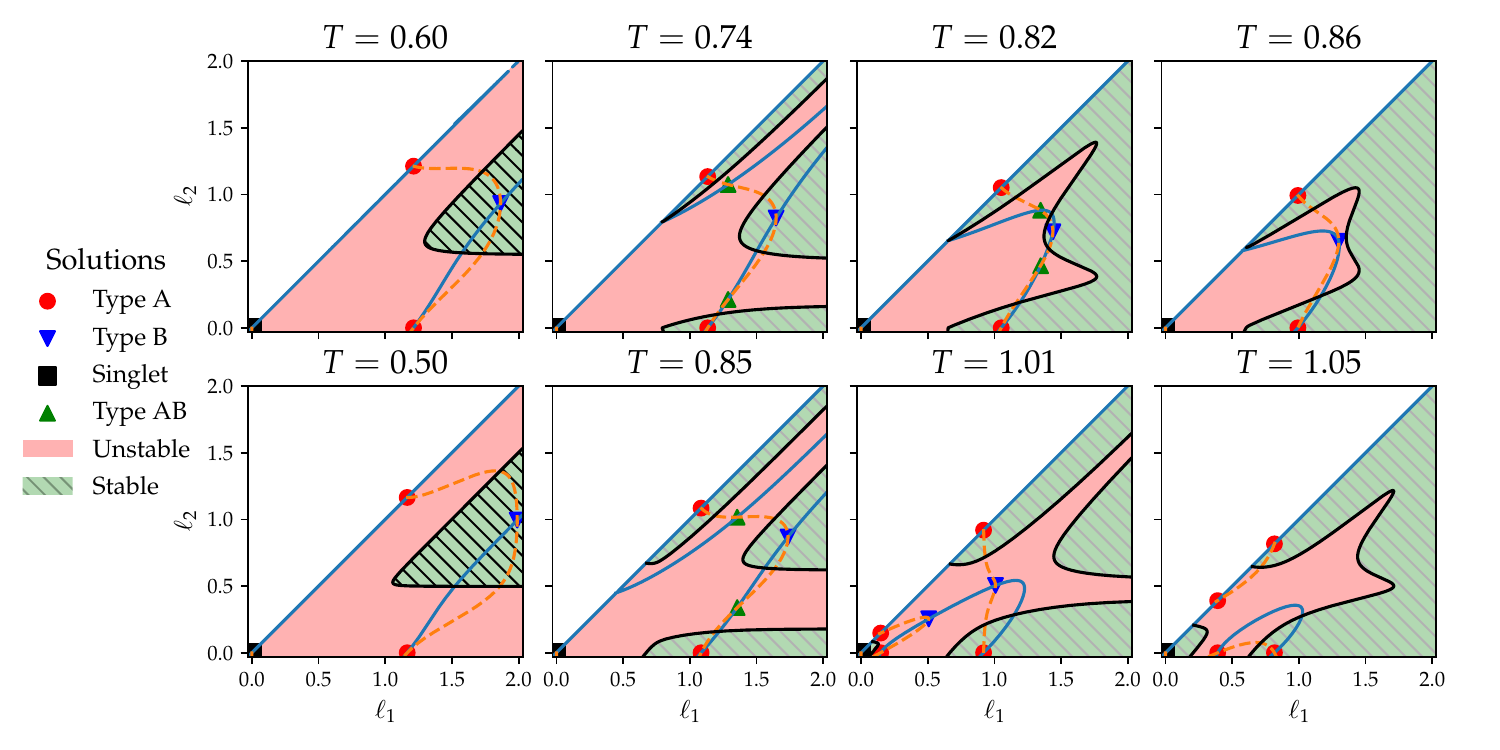}
  \end{center}
    \vskip -0.5cm
\caption{\small{Stable and unstable YT configurations, for $N\bb=\bb5$ (top four subplots) and $N\bb=\bb7$ (bottom four subplots) and various temperatures,
as denoted by the plots' titles. Temperatures are in units of $T_0$ and the lengths of YTs are in units of $n$.
 All plots correspond to $p_1\bb=\bb1$ and $p_2\bb=\bb N\bb-\bb 2$.
Black lines represent boundaries between locally stable and unstable configurations whereas blue and dashed orange lines represent the equilibrium equations of the system, their intersection marking the solutions.\\
For both $N\bb=\bb5$ and $N\bb=\bb7$, the singlet $\ell_1=\ell_2=0$ is stable only for $T >T_0$.\\
For $N\bb=\bb5$, the first plot corresponds to temperature $T \bb<\bb T_{\rm A}^{\rm (m)}$ where only the type B solution is stable. The second and third plots
correspond to temperatures $T_{\rm A}^{\rm (m)} < T < T_{\rm B}^{\rm (un)}$, where both type A and B solutions are locally stable. In this
temperature range, a pair of solutions of type AB emerge. These solutions are always unstable, and they represent YT configurations intermediate
between types A and B, as in \eqn{ABsol}, becoming type A as $T\to T_{\rm A}^{\rm (m)}$ and type B as
$T\to T_{\rm B}^{\rm (un)}$. The lower (upper) AB solution corresponds to the left  (right) YT in  \eqn{ABsol}.
Finally, the fourth plot corresponds to a temperature $T_{\rm B}^{\rm (un)} < T < T_0$, where the type A solution is stable
but the type B is not.\\
For ${N\bb=\bb7}$, the first plot corresponds to temperature $T \bb<\bb T_{\rm A}^{\rm (m)}$ where only the type B solution is stable.
The second plot corresponds to temperatures $T_{\rm A}^{\rm (m)} < T < T_{\rm B}^{\rm (un)}$, where both type A and B solutions are stable.
In this range a pair of solutions of type AB emerge, which are always unstable. The third plot corresponds to temperature $T_{\rm B}^{\rm (un)}\bb<\bb T_0 \bb<\bb T < T_{\rm B}^{(c)}$, where the type A solution and singlet are (meta)stable, but the type B one is not. Lastly, the fourth plot corresponds to temperatures $T_{\rm B}^{(c)} < T < T_{\rm A}^{(c)}$, where only the singlet and type A solutions exist and are (meta)stable, and some of the type A solutions are unstable.
}}
  \label{fig:contours_N7}
\end{figure}

\no
The simplest such example is solutions with  $p_1=1$ and $p_2=N-2$ with general values of $\a_i$. 
We present all different solutions for comparison, including A, B, the singlet, as well as these new solutions,
in fig. \ref{fig:contours_N7} for $N=5$ and $N=7$.
The new solutions start as  type A for $T=T_{\rm A}^{(\rm m)}$ and  then, as the temperature increases, they become 
an intermediate configuration between type A and type B.  
Finally, they become a type B configuration at $T=T_{\rm B}^{(\rm un)}$. We call these solutions AB and, schematically, they correspond to the following YT
\be
\label{ABsol}
{\rm AB}: \scalebox{.7}{ \begin{ytableau}
\none   &     \boldsymbol{ \bullet}   &     \boldsymbol{ \bullet}   & &  & & 
\end{ytableau}} \qquad {\rm or} \quad 
\scalebox{.7}{ \begin{ytableau}
\none   &  \boldsymbol{ \bullet}      &  \boldsymbol{ \bullet}     &  \boldsymbol{ \bullet}    &   \boldsymbol{ \bullet}    &   &  
\end{ytableau}} \ .
\ee
This may be thought of as resulting from the solution \eqn{at} by adding dots on the left YT and removing dots from the right YT.

\begin{figure}[!ht]
  \begin{center}
    \includegraphics[width=0.65\textwidth]{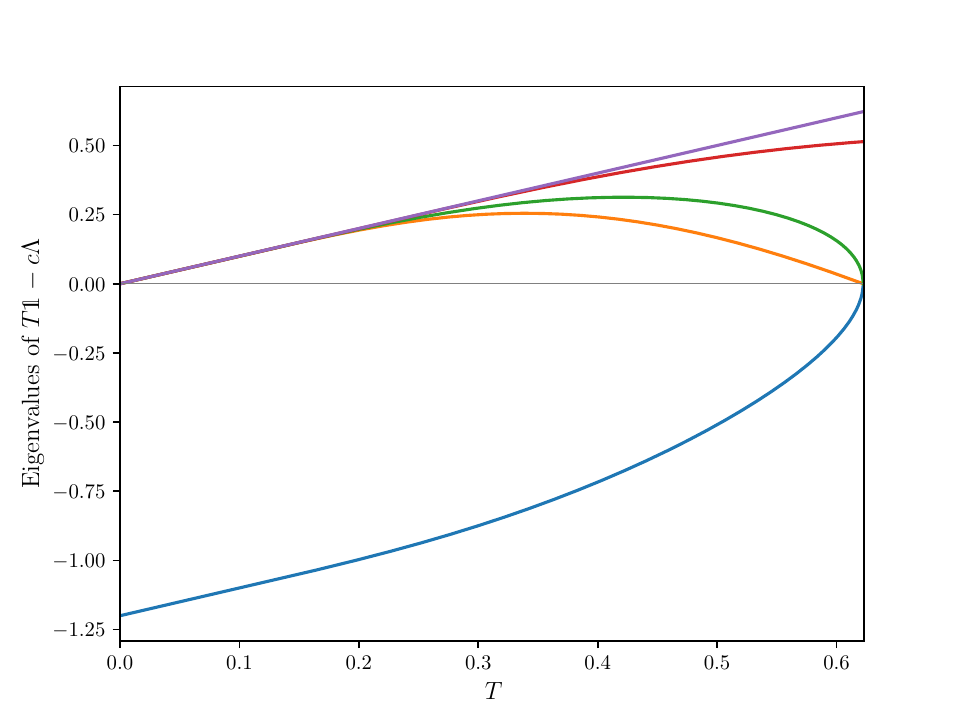}
  \end{center}
  \vskip -1 cm 
  \caption{\small{Eigenvalues of a generic solution, for $N=5$, $p_1=1$ and $p_2=2$, as a function of temperature, which is in units of $T_0$. This solution exists for temperatures $T \lesssim 0.622$. One eigenvalue is always negative and hence the solution is unstable.}
  \label{fig:eigenvalues_N5_asymmetric}}
\end{figure}

\no
Other solutions, also unstable, can be found for values of $(p_1,p_2)\neq (1,N-2)$. These solutions violate the self-conjugation symmetry.
For the simplest case with 
$(p_1,p_2)= (1,N-3)$ and all three $\a_i$ different, the stability matrix $T\, \mathbb{1} - c \L$ has negative eigenvalues for all temperatures.
We plotted these eigenvalues, in the indicative case $N=5$, as a function of the temperature in fig. \ref{fig:eigenvalues_N5_asymmetric}.
Such solutions corresponds to YTs of the general form \eqn{genn}. In particular, for the case with $(p_1,p_2)=(1,N-3)$ they assume the form 
\be
\scalebox{.7}{ \begin{ytableau}
		\none   & \boldsymbol{\bullet}   \\
		\none   & \boldsymbol{\bullet}
	\end{ytableau}}  \cdots \hskip -.4 cm \scalebox{.7}{ \begin{ytableau}
		\none   & \boldsymbol{\bullet}  & & & \\
		\none   & \boldsymbol{\bullet}
	\end{ytableau}} \cdots \hskip -.4 cm \scalebox{.7}{  \begin{ytableau}
		\none  & & &
	\end{ytableau}}\ .
\ee
Similar plots to fig. \ref{fig:eigenvalues_N5_asymmetric} can be produced for higher values of $N$, but we will not present them here.

\section{Some group theory results}
\label{cgroupth}

It is interesting that the parametrization of the ferromagnetic coupling constant $c$ in  
\eqn{cadj} in terms of the critical temperature $T_0$ below which the singlet becomes unstable can be written in the form 
\be
\label{cggrp}
c={T_0 \ov 2} {\dim G\ov C^{(2)}_\chi}\ ,
\ee
where $\dim G$ is the dimension of the group (here, $SU(N)$) and 
$C^{(2)}_\chi$ is the eigenvalue of the corresponding quadratic Casimir operator.
This expression was conjectured in \cite{PShigher} and was shown to work for the fundamental, symmetric, and antisymmetric irreps. 
Here we fill some details and show that \eqn{cggrp} also holds for the adjoint representation.
The quadratic Casimir operator is given by \cite{PScompo}
\be
\label{c2irrep}
C_2 ({\bf k}) = {1\over 2} \sum_{i=1}^N \left( k_i - {k\over N}\right)^{\bb 2} - {N(N^2 -1) \over 24}\ ,\quad k=\sum_{i=1}^N k_i\ .
\ee
The corresponding YT has $N-1$ rows with length given by 
\be
\ell_i = k_i - k_N+i-N\ ,\qq i=1,2,\dots, N-1\ .
\ee
The singlet corresponds to $k_i=N-i$,  $i=1,\dots ,N$.  For the 
fundamental representation we  increase $k_1$ by one, leaving the other $k_i$'s as they are.
For the symmetric representation we increase $k_1$ by two, whereas 
for the antisymmetric representation we increase $k_1$ as well as $k_2$ by one.
For the adjoint representation we increase $k_1$ by two and $k_i$, for $i=2,3,\dots N-1$, by one. 
Using the above, we find for the irreps we consider
\be
\label{c2irrepp}
\begin{split}
& 
C^{(2)}_{\rm fund} =  { N^2-1\ov 2 N}\ ,   \qquad C^{(2)}_{\rm sym} =  {(N-1)(N+2)\ov  N}\  ,
\\
&
C^{(2)}_{\rm ant} =  { (N-2)(N+1)\ov  N}\   ,\qquad  C^{(2)}_{\rm adj} =  {N}\ .
\end{split}
\ee
The expressions for these Casimirs are, of course, well known, but we wanted to fix the normalization and make contact with the $\bf k$ representation.
Clearly, for $G=SU(N)$ we have $\dim G=N^2-1$ and \eqn{cggrp} reduces to \eqn{cadj}.

\end{document}